\documentclass[preprintnumbers,article,amsmath,amssymb,floatfix,10pt,prd,twocolumn,
superscriptaddress,nofootinbib]{revtex4-2}
\usepackage{bm}
\usepackage{amsfonts}
\usepackage{latexsym}
\usepackage[latin1]{inputenc}
\usepackage{graphicx}
\usepackage{amsmath}
\usepackage{bigints} 
\usepackage{palatino}
\usepackage{mathpazo}
\usepackage{textcomp}
\usepackage{float}
\usepackage{booktabs}
\usepackage{dcolumn}
\usepackage{ragged2e}
\usepackage{hyperref}
\hypersetup{colorlinks,citecolor=blue}
\hypersetup{colorlinks=true,linkcolor=red,filecolor=magenta,    urlcolor=blue}
\usepackage{amsmath}
\usepackage{xcolor}
\usepackage{orcidlink}
\usepackage{epsfig}
\usepackage{subfigure}
\usepackage{commath}

\def\jnl@style{\it}
\def\aaref@jnl#1{{\jnl@style#1}}

\def\aaref@jnl#1{{\jnl@style#1}}

\def\aj{\aaref@jnl{AJ}}                   
\def\apj{\aaref@jnl{ApJ}}                 
\def\apjl{\aaref@jnl{ApJ}}                
\def\apjs{\aaref@jnl{ApJS}}               
\def\apss{\aaref@jnl{Ap\&SS}}             
\def\aap{\aaref@jnl{A\&A}}                
\def\aapr{\aaref@jnl{A\&A~Rev.}}          
\def\aaps{\aaref@jnl{A\&AS}}              
\def\mnras{\aaref@jnl{Mon.~Not.~Roy.~Astron.~Soc.}}             
\def\prd{\aaref@jnl{Phys.~Rev.~D}}        
\def\prc{\aaref@jnl{Phys.~Rev.~C}}  
\def\prl{\aaref@jnl{Phys.~Rev.~Lett.}}    
\def\qjras{\aaref@jnl{QJRAS}}             
\def\skytel{\aaref@jnl{S\&T}}             
\def\ssr{\aaref@jnl{Space~Sci.~Rev.}}     
\def\zap{\aaref@jnl{ZAp}}                 
\def\nat{\aaref@jnl{Nature}}              
\def\aplett{\aaref@jnl{Astrophys.~Lett.}} 
\def\apspr{\aaref@jnl{Astrophys.~Space~Phys.~Res.}} 
\def\physrep{\aaref@jnl{Phys.~Rep.}}      
\def\physscr{\aaref@jnl{Phys.~Scr}}       
\def\commat{\aaref@jnl{Comm.~Math.~Phys.}}              
\def\science{\aaref@jnl{Science}}               
\def\cqg{\aaref@jnl{Classical Quant.~Grav.}}            
\def\jpcs{\aaref@jnl{JPCS}}                                     
\def\ijmpd{\aaref@jnl{Int.~J.~Mod.~Phys.~D}}                    
\def\grg{\aaref@jnl{Gen.~Relat.~Gravit.}}               
\def\rpp{\aaref@jnl{Rep.~Prog.~Phys.}}          
\def\npa{\aaref@jnl{Nucl.~Phys.~A}}        
\def\lrr{\aaref@jnl{Living Rev.~Rel.}}                   
\def\jcap{\aaref@jnl{J.~Cosmology Astropart.~Phys.}}    
\def\rmp{\aaref@jnl{Rev.~Mod.~Phys.}}   
\def\epjc{\aaref@jnl{Eur.~Phys.~J.~C}} 
\def\plb{\aaref@jnl{~Phy.~Lett.~B}} 
\def\mpla{\aaref@jnl{Mod.~Phy.~Lett.~A}} 
\def\arxiv{\aaref@jnl{arxiv.org}}

\allowdisplaybreaks[1]

\begin{document}
\color{black} 
\title{Existence of Wormhole Solutions in $f(Q,\,T)$ Gravity under Non-commutative Geometries}

\author{Moreshwar Tayde\orcidlink{0000-0002-3110-3411}}
\email{moreshwartayde@gmail.com}
\affiliation{Department of Mathematics, Birla Institute of Technology and
Science-Pilani,\\ Hyderabad Campus, Hyderabad-500078, India.}

\author{Zinnat Hassan\orcidlink{0000-0002-6608-2075}}
\email{zinnathassan980@gmail.com}
\affiliation{Department of Mathematics, Birla Institute of Technology and
Science-Pilani,\\ Hyderabad Campus, Hyderabad-500078, India.}

\author{P.K. Sahoo\orcidlink{0000-0003-2130-8832}}
\email{pksahoo@hyderabad.bits-pilani.ac.in}
\affiliation{Department of Mathematics, Birla Institute of Technology and
Science-Pilani,\\ Hyderabad Campus, Hyderabad-500078, India.}
\affiliation{Faculty of Mathematics \& Computer Science, Transilvania University of Brasov, Eroilor 29, Brasov, Romania}

%
\date{\today}

\begin{abstract}
In this paper, we have systematically discussed the existence of the spherically symmetric wormhole solutions in the framework of $f(Q,\,T)$ gravity under two interesting non-commutative geometries such as Gaussian and Lorentzian distributions of the string theory. Also, to find the solutions, we consider two $f(Q,\,T)$ models such as linear $f(Q,\,T)=\alpha\,Q+\beta\,T$ and non-linear $f(Q,\,T)=Q+\lambda\,Q^2+\eta\,T$ models in our study. We obtained analytic and numerical solutions for the above models in the presence of both non-commutative distributions. We discussed wormhole solutions analytically for the first model and numerically for the second model and graphically showed their behaviors with the appropriate choice of free parameters. We noticed that the obtained shape function is compatible with the flare-out conditions under asymptotic background. Further, we checked energy conditions at the wormhole throat with throat radius $r_0$ and found that NEC is violated for both models under non-commutative background. At last, we examine the gravitational lensing phenomenon for the precise wormhole model and determine that the deflection angle diverges at the wormhole throat.
\end{abstract}

\maketitle



\section{Introduction}\label{sec1}
It is well-known that black holes and wormholes are two effective solutions to the Einstein field equations. Einstein first estimated the presence of a black hole by his theory of General Relativity (GR) \cite{Rosen11}. A black hole is considered an area of space with a strong gravitational pull that the fastest-moving objects, even light, cannot escape. The existence of black holes has been recently investigated in \cite{B1,B2,B3,B4}; however, wormhole existence is still an open question. A discussion on this topic was presented in \cite{B5}. A study of past and current efforts to search for wormholes in the Universe was reported in \cite{B55}. For more interesting articles on this particular topic, readers can check some Refs. \cite{B6,B7,B8,B9,B10,B11,B12}.\\
Wormholes are generally tunnels with two ends each at different points in space-time, which was first suspected by Weyl \cite{Coleman} and subsequently by Wheeler \cite{Wheeler}. In a more concrete physical definition, it is basically some hypothetical topological characteristic of space-time that may operate as a shortcut via space-time topology. These are supported by matter that violates all the energy conditions at the throat in GR. After discovering Schwarzschild's black hole solutions, Flamm \cite{Flamm} first suggested the idea of a wormhole. Later in 1935, Einstein and Rosen used Flamm's concept and made a hypothetical bridge called the Einstein-Rosen bridge \cite{Einstein}. Nevertheless, the attraction to traversable wormholes has been growing interest after the prominent study of Morris, Thorne, and Yurtsever \cite{Thorne1, Thorne2}. Although it is a challenging task to construct wormhole solutions with ordinary matter, however, this issue has been determined for rotating cylindrical wormholes in the context of GR, and wormhole solutions respecting weak energy conditions (WEC) have been analyzed in \cite{Bronnikov1, Bronnikov2}. Moreover, it is known that wormholes can also exist with normal matter in modified theories of gravity \cite{Bhawal, Kunz, Montelongo, Mak, SenGupta, Galiakhmetov, Moraes1, Moraes2, Shaikh, Sakalh,R1,R10}.\\
In recent years, one could see substantial growth in the area of modified theories of gravity \cite{Laurentis}. The first extension of GR is the $f(R)$ gravity, which modifies the Einstein-Hilbert action by replacing the Ricci scalar with a function of the scalar curvature \cite{Hans A. Buchdahl,A. A. Starobinsky,A. D. Felice}. This modification changes the equations governing the gravitational field and can impact the behavior of gravity on various scales. The $f(R)$ gravity theory has been proposed as an explanation for the observed acceleration of the Universe's expansion, provides insights into the constraints on primordial inflation, and offers an alternative to the requirement of dark matter for explaining stellar dynamics, galaxy rotation curves, and galaxy morphology \cite{Capozziello/2012}. Lobo \textit{et al.} \cite{ Oliveira3} constructed traversable wormhole geometries in the context of $f(R)$ gravity by considering specific shape functions and several equations of state. Later, in the extension of $f(R)$ gravity, Shamir \textit{et al.} \cite{R3} explored the traversable wormhole solution in $f(R, G)$ gravity and Banerjee \textit{et al.} \cite{Ayan Banerjee} studied the wormhole solution satisfying the null energy condition with isotropic pressure in $f(R, T)$ gravity. Also, one can explore the Refs. \cite{R2,R4,R5,R6,R7,R8,R9,Tayde 1, Tayde 2} for more exciting articles on the modified theories of gravity. In the late '90s, the non-metricity theory came to light among researchers after the proposal of Symmetric Teleparallel Gravity \cite{Nester, Kalay, Conroy}. It is the torsion-less and curvature-less theory; consequently, gravitation is connected to the non-metricity tensor and affiliated with the nonmetricity scalar. In 2018, Jimenez et al. generalized this theory named $f(Q)$ gravity \cite{Jimenez}, where the gravitational field is expressed by the non-metricity scalar $Q$ only. This theory has successfully encountered various perturbation and observational data such as the Supernovae type Ia (SNIa), Baryonic Acoustic Oscillations (BAO), Cosmic Microwave Background (CMB), Redshift Space Distortion (RSD), etc., \cite{Soudi, Banos, Salzano, Koivisto11}. Moreover, in Ref. \cite{Anagnostopoulos}, the authors provide evidence that the non-metricity $f(Q)$ gravity could challenge $\Lambda$CDM. In addition,  $f(Q)$ gravity has been successfully implemented in the field of astrophysical objects. In \cite{Fell}, Black holes in $f(Q)$ gravity have been investigated. The application of the spherically symmetric configurations in f(Q) gravity has been discussed in \cite{Zhai}. In \cite{Hassan11}, Hassan et al. have analyzed wormhole geometries in $f(Q)$ gravity by choosing different linear equations of state under linear and non-linear models. They obtained the exact solutions for the linear model, and using the volume integral quantifier tool, they confirmed that a small amount of exotic matter is needed for a traversable wormhole \cite{Hassan11}. Further, Mustafa et al. have acquired wormhole solutions from the Karmarkar condition in $f(Q)$ gravity, providing the prospect of developing wormholes satisfying energy conditions \cite{Mustafa}. For furthermore applications of $f(Q)$ gravity, readers can check some Refs. \cite{Hohmann,Koivisto,De,Kuhn,Zhao,Ghosh,Wang2,Calza,Pradhan1}.\\
Recently, a matter-geometry coupling in the form of $f(Q,\,T)$ theories were presented \cite{Y.Xu} in which any viable function of the non-metricity scalar $Q$ and the trace of the energy-momentum tensor $T$ represented the Lagrangian. Although $f(Q,\,T)$ is a newly proposed gravity, it has many useful applications in the cosmological aspects. The first cosmological significances in $f (Q,\,T )$ gravity have been discussed in \cite{Y.Xu}, while in \cite{Arora111}, the late-time accelerated expansion with observational constraints was studied for the $f(Q,\,T)$ gravity model. Later, Cosmological inflation \cite{Shiravand},  Baryogenesis \cite{Bhattacharjee}, Cosmological perturbations \cite{Najera}, and Reconstruction of $f (Q,\,T)$ Lagrangian \cite{Gadbail} have been widely investigated. However, we could see less study in the astrophysical scenario in this modified gravity. In \cite{Tayde 3}, the authors investigated static spherically symmetric wormhole solutions in $f(Q,\,T)$ gravity for linear and non-linear models under the different equations of state relations. They confirm that the exact solutions could be found for the linear model, whereas, for the non-linear model, it is almost impossible to obtain analytically. Nevertheless, except for the aforementioned article, no more studies were conducted on this gravity theory which is still in its infancy at best. This motivated us to investigate wormhole solutions under non-commutative geometries in this $f(Q,\,T)$ gravity.\\
The goal of noncommutative geometry is to create a unified framework where one may consider the spacetime gravitational forces as a combined form of strong and weak forces with gravity. Non-commutativity has the critical attribute of replacing point-like structures with smeared objects, and hence it relates to spacetime discretization and the commutator described by $\left[x^\alpha,\,x^\beta\right]=i\,\Theta^{\alpha\beta}$ \cite{Doplicher, Spallucci, Nicolini}, where $\Theta^{\alpha\beta}$ represented as a second-order anti-symmetric matrix of dimension $(\text{length})^2$. It is analogous to the discretization of phase space by the Planck constant $\hbar$ \cite{Smailagic1}. The standard concept of mass density in the form of the Dirac delta function does not hold true in non-commutative space. Thus, instead of using the Dirac delta function, Gaussian and Lorentzian distributions of negligible length $\sqrt{\Theta}$ can be used to illustrate this distribution impact. The static, spherically symmetric smeared, and particle-like gravitational source handles the geometry of Gaussian distribution with non-commutative nature with the maximum mass M retains. Whereas for Lorentzian distribution, the density capacity of the particle-like mass M can be assumed. In this case, the entire mass M can be regarded as a form of the diffused unified object, and $\Theta$ is the non-commutative parameter. This kind of matter distribution was used by Nicolini and Spalluci \cite{Nicolini2} to specify the physical substances of short-separated divergences of non-commutating coordinates in probing black holes. In \cite{Islam}, the authors investigated wormhole solutions for higher-dimension under Gaussian distributions and noticed that wormhole solutions could exist in the four and five dimensions only. Also, Lorentzian distributed non-commutative wormholes in the framework of $f(R)$ gravity have been studied in \cite{Banerjee}. Recently, wormhole solutions have been investigated for both distributions in $f(Q)$ gravity \cite{Sokoliuk, Hassan1} as well as other modified theories of gravity \cite{Shi-Qin Song, Nilofar Rahman, Yihu Feng, Anshuman Baruah, M. Sharif 1, Farook Rahaman 1, Mubasher Jamil, Shamaila Rani}.\\
Gravitational lensing is one of the initial applications of GR ever investigated \cite{Einstein10}. It occurs when an astronomical object is enormous enough to bend the falling light into a lens, allowing the observer to gather more data regarding the source than feasible. It has grown its interest among researchers after the observational verification of the theoretical prediction of light bending \cite{Dyson,Eddington}. It not only bound us to study stars and galaxies but also enabled us to investigate extrasolar planets, dark energy, and dark matter. Recently, a group of researchers successfully made the first detection of astrometric microlensing to calculate the mass of the white dwarf Stein 2051 B \cite{Sahu}. A charming characteristic of gravitational lensing is that, in unstable light rings, light can undergo an unboundedly large (theoretically infinite) amount of bending \cite{Bozza1,Virbhadra,Bozza2}. As a result of such strong gravitational lensing, a large number of relativistic images are formed. Also, it is well-known that strong and weak gravitational lensing is a powerful tool to effective phenomenon tool for analyzing gravitational fields around different astrophysical objects such as black holes and wormholes. One can check some works of literature where wormholes were inspected by gravitational lensing on a large scale in theoretical physics as well as astrophysics \cite{A1,A2,A3,A4,A5,A6,A7,A8,A9,A10}. \\
This article is organized as follows: We have introduced the basic formalism of $f(Q,\,T)$ gravity in Sec. \ref{sec2} and obtained the corresponding field equations. A brief review of the non-commutative geometry with linear form of $f(Q,\,T)$ is presented in  Sec. \ref{sec4} and with a non-linear form of $f(Q,\,T)$ is presented in  Sec. \ref{sec5}. To examine the deflection angle at the wormhole throat, Sec. \ref{sec6} will be dedicated to analyzing the gravitational lensing phenomenon for the specific model of Sec. \ref{sec4}. Finally, we conclude our results in the last section \ref{sec7}.

\section{Traversability conditions of wormhole and field equations in $f(Q,\,T)$ gravity}
\label{sec2}
We consider the spherically symmetric and static wormhole metric in Schwarzschild coordinates $(t,\,r,\,\theta,\,\Phi)$ is given by \cite{Visser, Morris}
\begin{equation}\label{10}
ds^2=e^{2\phi(r)}dt^2-\left(1-\frac{b(r)}{r}\right)^{-1}dr^2-r^2\,d\theta^2-r^2\,\sin^2\theta\,d\Phi^2\,,
\end{equation}
where $b(r)$ denotes the shape function, and it is used to define the shape of the wormholes. The function $\phi(r)$ represents the redshift function related to the gravitational redshift. Also, to have a wormhole to be traversable, the shape function $b(r)$ should meet the flaring-out condition, given by $(b-b'r)/b^2>0$ \cite{Morris} and at the wormhole throat $b(r_0)=r_0$, the condition $b^{\,\prime}(r_0)<1$ is imposed ($r_0$ denotes the throat radius). Additionally, the asymptotic flatness condition, that is, the limit $\frac{b(r)}{r}\rightarrow 0$ as $r\rightarrow \infty$, is also required. Besides, $\phi(r)$ must be finite everywhere to avoid the event horizon. In Einstein's GR, fulfilling the above criteria may ensure the existence of exotic matter at the throat of the wormhole.\\
Now, we are going to briefly present some generalities about the $f(Q,\,T)$ gravity.
We consider the action for symmetric teleparallel gravity proposed in \cite{Y.Xu}
\begin{equation}\label{1}
\mathcal{S}=\int\frac{1}{16\pi}\,f(Q,\,T)\sqrt{-g}\,d^4x+\int \mathcal{L}_m\,\sqrt{-g}\,d^4x\, .
\end{equation}
The function of non-metricity $Q$ and a trace of the energy-momentum tensor $T$ is represented as $f(Q,\,T)$, where $g$ stands for the determinant of the metric $g_{\mu\nu}$, and $\mathcal{L}_m$ is the matter Lagrangian density.\\
The non-metricity tensor is defined by the equation  \cite{Jimenez}\\
\begin{equation}\label{2}
Q_{\lambda\mu\nu}=\bigtriangledown_{\lambda} g_{\mu\nu}\,.
\end{equation}
Also, the non-metricity conjugate or superpotential can be defined as
\begin{equation}\label{3}
P^\alpha\;_{\mu\nu}=\frac{1}{4}\left[-Q^\alpha\;_{\mu\nu}+2Q_{(\mu}\;^\alpha\;_{\nu)}+Q^\alpha g_{\mu\nu}-\tilde{Q}^\alpha g_{\mu\nu}-\delta^\alpha_{(\mu}Q_{\nu)}\right].
\end{equation}
And, two traces of the non-metricity tensor are given by
\begin{equation}
\label{4}
Q_{\alpha}=Q_{\alpha}\;^{\mu}\;_{\mu},\; \tilde{Q}_\alpha=Q^\mu\;_{\alpha\mu}.
\end{equation}
The non-metricity scalar is represented as \cite{Jimenez}
\begin{eqnarray}
\label{5}
Q &=& -Q_{\alpha\mu\nu}\,P^{\alpha\mu\nu}\\
&=& -g^{\mu\nu}\left(L^\beta_{\,\,\,\alpha\mu}\,L^\alpha_{\,\,\,\nu\beta}-L^\beta_{\,\,\,\alpha\beta}\,L^\alpha_{\,\,\,\mu\nu}\right),
\end{eqnarray}
 where $L^\beta_{\,\,\,\mu\nu}$ represents the disformation defined by 
\begin{equation}\label{6}
L^\beta_{\,\,\,\mu\nu}=\frac{1}{2}Q^\beta_{\,\,\,\mu\nu}-Q_{(\mu\,\,\,\,\,\,\nu)}^{\,\,\,\,\,\,\beta}.
\end{equation}

Now, the gravitational equations of motion can be obtained by varying the action with respect to the metric tensor $g_{\mu\nu}$ and are represented as
\begin{multline}\label{7}
\frac{-2}{\sqrt{-g}}\bigtriangledown_\alpha\left(\sqrt{-g}\,f_Q\,P^\alpha\;_{\mu\nu}\right)-\frac{1}{2}g_{\mu\nu}f + f_T \left(T_{\mu\nu} +\Psi_{\mu\nu}\right) \\
-f_Q\left(P_{\mu\alpha\beta}\,Q_\nu\;^{\alpha\beta}-2\,Q^
{\alpha\beta}\,\,_{\mu}\,P_{\alpha\beta\nu}\right)=8\pi T_{\mu\nu},
\end{multline}

where $f_Q=\frac{\partial f}{\partial Q}$ and $f_T=\frac{\partial f}{\partial T}$.

The energy-momentum tensor for the fluid depiction of spacetime can be described as
\begin{equation}\label{8}
T_{\mu\nu}=-\frac{2}{\sqrt{-g}}\frac{\delta\left(\sqrt{-g}\,\mathcal{L}_m\right)}{\delta g^{\mu\nu}},
\end{equation}
and
\begin{equation}\label{9}
\Psi_{\mu\nu}=g^{\alpha\beta}\frac{\delta T_{\alpha\beta}}{\delta g^{\mu\nu}}.
\end{equation}
In this study, we analyze wormhole solutions assuming an anisotropic energy-momentum tensor which is given by \cite{Morris, Visser} and is represented by Eq. \eqref{11} as follows:
\begin{equation}\label{11}
T_{\mu}^{\nu}=\left(\rho+p_t\right)u_{\mu}\,u^{\nu}-p_t\,\delta_{\mu}^{\nu}+\left(p_r-p_t\right)v_{\mu}\,v^{\nu},
\end{equation}
where $\rho$ indicates the energy density, $u_{\mu}$ and $v_{\mu}$ are the four-velocity vector and unitary space-like vectors, respectively. Both vectors satisfy the conditions $u_{\mu}u^{\nu}=-v_{\mu}v^{\nu}=1$, and $p_r$ and $p_t$ represent the radial and tangential pressures, respectively. Both are functions of the radial coordinate $r$, and the trace of the energy-momentum tensor is given by $T=\rho-p_r-2p_t$.\\
We use the matter Lagrangian $\mathcal{L}_m=-P$ \cite{Correa} in this article, which leads to the following form for Eq. \eqref{9}:
\begin{equation}\label{12}
\Psi_{\mu\nu}=-g_{\mu\nu}\,P-2\,T_{\mu\nu},
\end{equation}
where $P$ is the total pressure, which can be expressed as $P=\frac{p_r+2\,p_t}{3}$.

The non-metricity scalar $Q$ for the metric \eqref{10} is given by \cite{Tayde 3} as follows:
\begin{equation}\label{13}
Q=-\frac{b}{r^2}\left[\frac{rb^{'}-b}{r(r-b)}+2\phi^{'}\right].
\end{equation}\\
The corresponding field equations for $f(Q,\,T)$ gravity are presented below \cite{Tayde 3}
\begin{multline}\label{14}
8 \pi  \rho =\frac{(r-b)}{2 r^3} \left[f_Q \left(\frac{(2 r-b) \left(r b'-b\right)}{(r-b)^2}+\frac{b \left(2 r \phi '+2\right)}{r-b}\right)
\right. \\ \left.
+\frac{2 b r f_{\text{QQ}} Q'}{r-b}+\frac{f r^3}{r-b}-\frac{2r^3 f_T (P+\rho )}{(r-b)}\right],
\end{multline}
\begin{multline}\label{15}
8 \pi  p_r=-\frac{(r-b)}{2 r^3} \left[f_Q \left(\frac{b }{r-b}\left(\frac{r b'-b}{r-b}+2 r \phi '+2\right)
\right.\right. \\ \left.\left.
-4 r \phi '\right)+\frac{2 b r f_{\text{QQ}} Q'}{r-b}+\frac{f r^3}{r-b}-\frac{2r^3 f_T \left(P-p_r\right)}{(r-b)}\right],
\end{multline}
\begin{multline}\label{16}
8 \pi  p_t=-\frac{(r-b)}{4 r^2} \left[f_Q \left(\frac{\left(r b'-b\right) \left(\frac{2 r}{r-b}+2 r \phi '\right)}{r (r-b)}+
\right.\right. \\ \left.\left.
\frac{4 (2 b-r) \phi '}{r-b}-4 r \left(\phi '\right)^2-4 r \phi ''\right)
-4 r f_{\text{QQ}} Q' \phi '\right.\\\left.
+\frac{2 f r^2}{r-b}-\frac{4r^2 f_T \left(P-p_t\right)}{(r-b)}\right].
\end{multline}
By utilizing these particular field equations, various wormhole solutions can be examined within the framework of $f(Q,\,T)$ gravity models.

\subsection{Energy conditions}
The classical energy conditions, developed from the Raychaudhuri equations, are used to discuss physically realistic matter configurations. The four energy conditions that are commonly used are the null energy condition (NEC), weak energy condition (WEC), dominant energy condition (DEC), and strong energy condition (SEC). The NEC is considered to be the most significant of these conditions for wormhole solutions in general relativity because it is associated with the energy density required to keep the wormhole throat open. Violation of the NEC in the vicinity of the throat of a wormhole would imply the presence of exotic matter with negative energy density, which is not found in standard matter sources. The geometrical connection between attractive gravity and the Raychaudhuri equation \cite{S. Carroll} provides the basis for this theory and is given by
\begin{equation}\label{16a}
\frac{d\theta}{d\tau}-\omega_{\mu\nu}\,\omega^{\mu\nu}+\sigma_{\mu\nu}\sigma^{\mu\nu}+\frac{1}{3}\theta^2+R_{\mu\nu}u^\mu\,u^\nu=0\,,
\end{equation}
\begin{equation}\label{16b}
\frac{d\theta}{d\tau}-\omega_{\mu\nu}\,\omega^{\mu\nu}+\sigma_{\mu\nu}\sigma^{\mu\nu}+\frac{1}{2}\theta^2+R_{\mu\nu}v^\mu v^\nu=0\,,
\end{equation}
where $\sigma^{\mu\nu}$ and $\omega_{\mu\nu}$ are the shear and the rotation associated with the vector field $u^\mu$ respectively; whereas $\theta$ is the expansion factor and $\tau$ is the positive parameter. For attractive nature of gravity ($\theta<0$) and neglecting the quadratic terms, the Raychaudhuri equations \eqref{16a} and \eqref{16b} satisfy the following conditions
\begin{equation}
R_{\mu\nu}u^\mu\,u^\nu\geq0\,,
\end{equation}
\begin{equation}
R_{\mu\nu} v^\mu v^\nu\geq0\,.
\end{equation}
The concept of energy conditions plays a crucial role in the study of gravity and the behavior of matter in the universe. Energy conditions provide a set of constraints on the stress-energy tensor, which describes the distribution of matter and energy in spacetime. They can be expressed as follows:\\
$\bullet$ The weak energy condition (\textbf{WEC}) requires that the energy density $\rho$ is always non-negative, and the sum of the energy density and pressure in any direction is also non-negative, i.e.,
$\rho\geq0$,\,\, $\rho+p_r\geq0$,\,\, and \,\, $\rho+p_t\geq0$.\\
$\bullet$ The null energy condition (\textbf{NEC}) is a stronger condition than WEC and requires that the sum of the energy density and pressure in all directions is non-negative, i.e., $\rho+p_r\geq0$\,\, and \,\, $\rho+p_t\geq0$.\\
$\bullet$ The dominant energy condition (\textbf{\textbf{DEC}}) requires that the energy density is non-negative, and the pressure in any direction is dominated by the energy density, i.e.,  $\rho\geq0$,\,\, $\rho+p_r\geq0$,\,\, $\rho+p_t\geq0$,\,\, $\rho-p_r\geq0$,\,\, and \,\, $\rho-p_t\geq0$.\\
$\bullet$ The strong energy condition (\textbf{SEC}) is the strongest energy condition and requires that the energy density and the sum of the energy density and pressure in all directions are non-negative, i.e.,
 $\rho+p_r\geq0$,\,\, $\rho+p_t\geq0$,\,\, and \,\, $\rho+p_r+2p_t\geq0$.\\
In summary, energy conditions provide powerful constraints on the behavior of matter in the universe and play a crucial role in the study of wormholes.

\section{Wormhole solutions with $f(Q,\,T)=\alpha\,Q+\beta\,T$}
\label{sec4}
In this section, we shall consider a specific and interesting $f(Q,\,T)$ model given by
\begin{equation}
\label{17}
f(Q,\,T)=\alpha\,Q+\beta\,T\,,
\end{equation}
where $\alpha$ and $\beta$ are dimensionless model parameters. Xu et al. \cite{Y.Xu} introduced this model, and it naturally describes an exponentially expanding Universe, with $\rho \propto e^{-H_0 t}$ \cite{Y.Xu}. Also, this model has been used to constrained by observational data of the Hubble parameter in \cite{Arora11}. With the same model, Loo et al. \cite{Loo1} investigated Bianchi type-I cosmology with observational (Hubble and Type Ia supernovae) datasets. Moreover, wormhole solutions have been studied for this model with the different equation of state (EoS) relations in \cite{Tayde 3}. In this study, we will check the strength of this model under non-commutative geometries. Using the above linear functional form \eqref{17} with constant redshift function, the field equations \eqref{14}-\eqref{16} can be read as
 \begin{equation}\label{18}
 \rho =\frac{\alpha  (12 \pi -\beta ) b'}{3 (4 \pi -\beta ) (\beta +8 \pi ) r^2}\,,
 \end{equation}
 \begin{equation}\label{19}
 p_r=-\frac{\alpha  \left(2 \beta  r b'-3 \beta  b+12 \pi  b\right)}{3 (4 \pi -\beta ) (\beta +8 \pi ) r^3}\,,
 \end{equation}
 \begin{equation}\label{20}
 p_t=-\frac{\alpha  \left((\beta +12 \pi ) r b'+3 b (\beta -4 \pi )\right)}{6 (4 \pi -\beta ) (\beta +8 \pi ) r^3}\,.
 \end{equation}
 In the following subsections, we will study wormhole solutions under Gaussian and Lorentzian distributions. We will investigate the effect of these non-commutative geometries on wormhole solutions through the behaviors of shape functions and energy conditions.
 
 \subsection{Gaussian distribution}
 \label{subsubsec1}
\begin{figure*}[h]
\centering
\includegraphics[width=14.5cm,height=6cm]{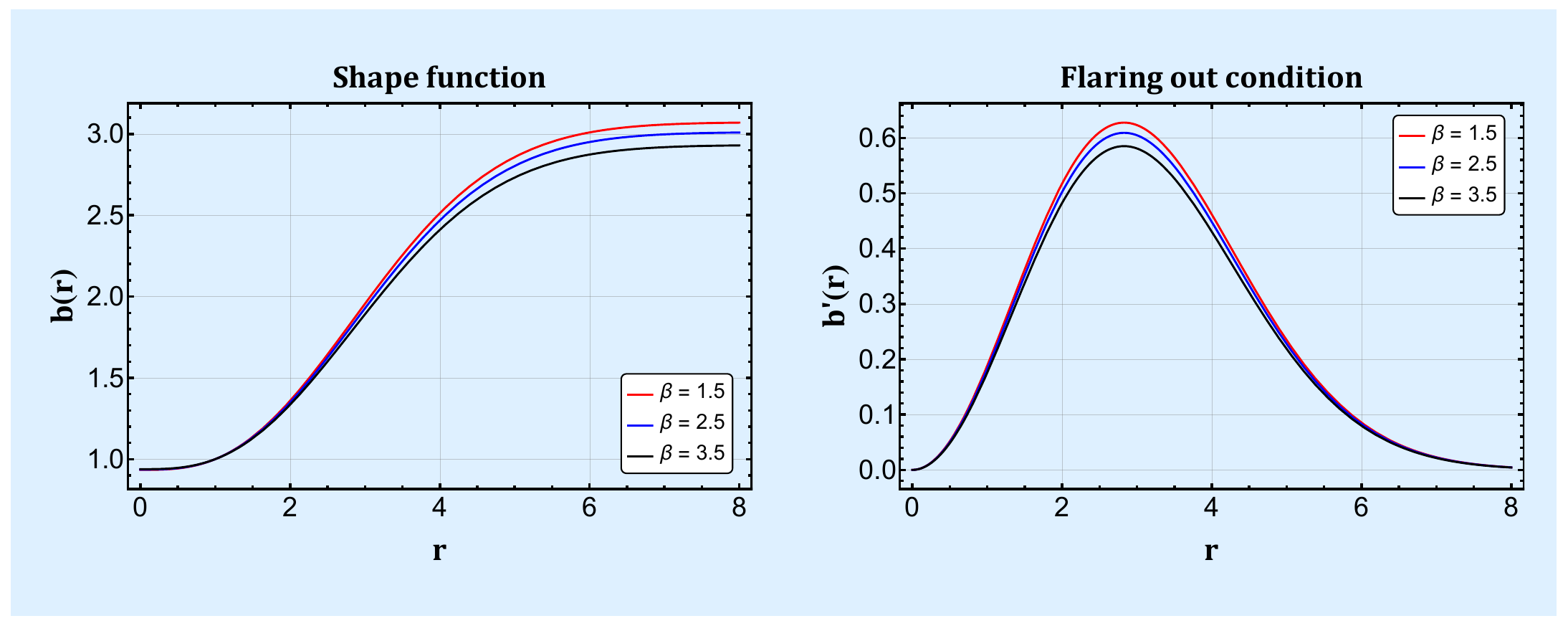}
\caption{Figure represents the behavior of shape function (\textit{left}) and flare-out condition (\textit{right}) against radial coordinate $r$ for different values of $\beta$. Also, we consider $\alpha=1,\, \Theta=2,\, M=1.1,\, \text{and} \, r_0 = 1$. It is clear that $b(r)$ shows a positively increasing behavior and $b'(r)$ satisfies the flaring out condition at the throat. See the text for details.}
\label{fig1}
\end{figure*}

\begin{figure*}[h]
\centering
\includegraphics[width=14.5cm,height=6cm]{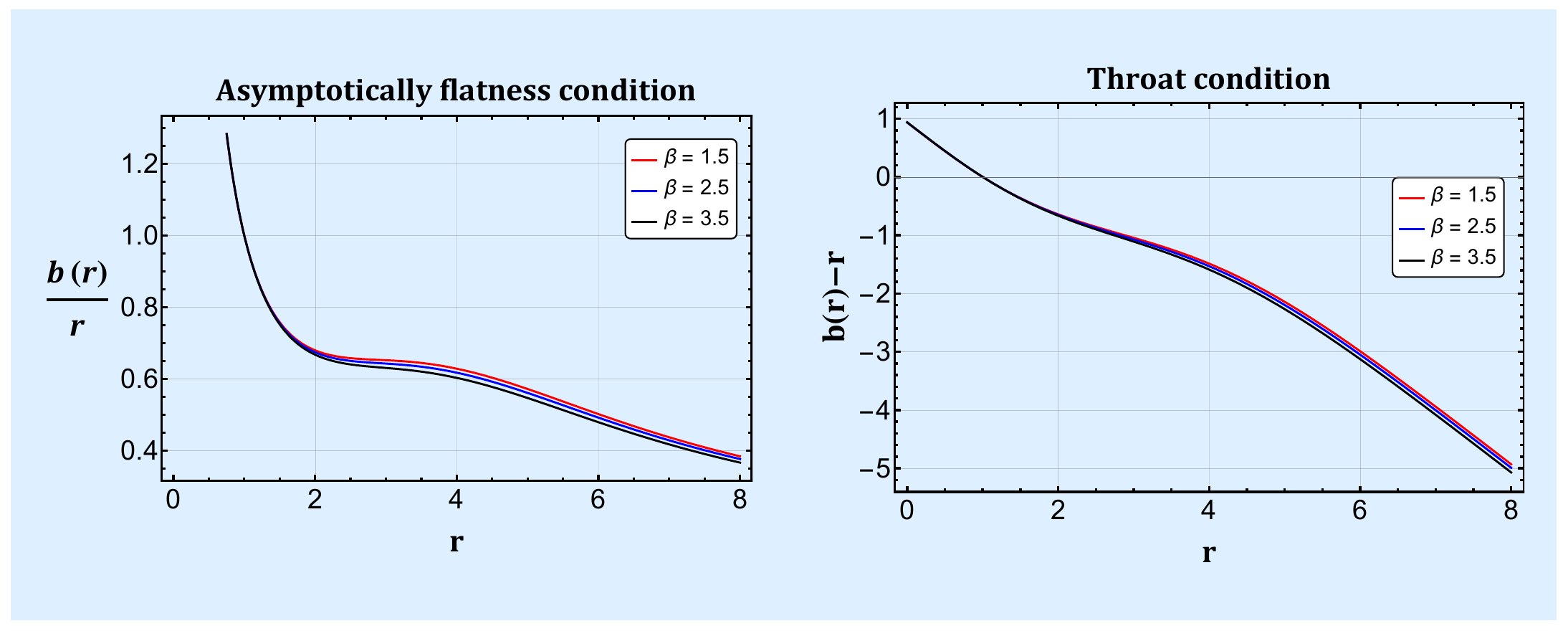}
\caption{Figure represents the behavior of asymptotically flatness condition (\textit{left}) and throat condition (\textit{right}) against radial coordinate $r$ for different values of $\beta$. Also, we consider $\alpha=1,\, \Theta=2,\, M=1.1,\, \text{and} \, r_0 = 1$. It is clear that $\frac{b(r)}{r}\rightarrow 0$ as $r\rightarrow \infty$ and $b(r)-r$  satisfies the throat condition. See the text for details.}
\label{fig2}
\end{figure*}

Here we consider the mass density of a static, spherically symmetric, smeared, particle-like gravitational source given by \cite{P. Nicoloni, A. Smailagic1}
 \begin{equation}\label{21}
 \rho =\frac{M e^{-\frac{r^2}{4 \Theta }}}{8 \pi ^{3/2} \Theta ^{3/2}}\,.
 \end{equation}
 The particle mass $M$, rather than being splendidly restricted at the point, diffused on a region of the direct estimate $\sqrt{\Theta}$.
This is because the uncertainty is encoded in the coordinate commutator.\\
 By comparing the Eqs. \eqref{18} and \eqref{21} , the differential equation under Gaussian distribution is given by 
 \begin{equation}\label{23}
 \frac{\alpha  (12 \pi -\beta ) b'(r)}{3 (4 \pi -\beta ) (\beta +8 \pi ) r^2}=\frac{M e^{-\frac{r^2}{4 \Theta }}}{8 \pi ^{3/2} \Theta ^{3/2}}\,.
\end{equation}
 By solving the above expressions, we obtain
 \begin{equation}\label{a1}
    b(r)=c_1+\mathcal{K}_1 \left(2 \sqrt{\pi } \Theta ^{3/2} \text{erf}\left(\frac{r}{2 \sqrt{\Theta }}\right)-2 \Theta  r e^{-\frac{r^2}{4 \Theta }}\right)\,,
 \end{equation}
 where $\mathcal{K}_1=\frac{3 \left(-\beta ^2-4 \pi  \beta +32 \pi ^2\right) M}{8 \pi ^{3/2} \alpha  (12 \pi -\beta ) \Theta ^{3/2}}$ and $c_1$ is the integrating constant. ``erf" is an error function and it can be defined by $\text{erf}(\Theta)=\frac{2}{\sqrt{\pi}}\bigintsss\limits_{0}^{\Theta}e^{-t^2}dt$. Now, to calculate $c_1$, we use the throat condition $b(r_0)=r_0$ in the above expression \eqref{a1}, we obtain
 \begin{equation}\label{a2}
    c_1=r_0-\mathcal{K}_1 \left(2 \sqrt{\pi } \Theta ^{3/2} \text{erf}\left(\frac{r_0}{2 \sqrt{\Theta }}\right)-2 \Theta  r_0 e^{-\frac{r_0^2}{4 \Theta }}\right),
 \end{equation}
and hence the shape function \eqref{a1} can be read as
 \begin{multline}\label{a3}
    b(r)=r_0+2\,\Theta \mathcal{K}_1\left[e^{-\frac{r^2+r_0^2}{4 \Theta }} \left(e^{\frac{r^2}{4 \Theta }} \left(\sqrt{\pi } \sqrt{\Theta } e^{\frac{r_0^2}{4 \Theta }} \left(\text{erf}\left(\frac{r}{2 \sqrt{\Theta }}\right)
    \right.\right.\right.\right. \\ \left.\left.\left.\left.
    -\text{erf}\left(\frac{r_0}{2 \sqrt{\Theta }}\right)\right)+r_0\right)-r e^{\frac{r_0^2}{4 \Theta }}\right)\right].
 \end{multline}
Now we will describe the graphical representation of the acquired shape function and the prerequisites for the wormhole's existence. Also, to archive it, we set appropriate choices for the concerned free parameters. First, we examine how the shape functions for the Gaussian distribution behave when the redshift function is assumed to be constant. The behavior of the shape function and flaring out condition for different values of $\beta$ were shown in Fig. \ref{fig1}. One can notice that the shape function exhibits favorably increasing behavior. However, by increasing the value of the model parameter $\beta$, the shape function shows decreasing behavior. Also, the right graph depicted that the flaring out condition $b'(r_0)<1$ is satisfied at the wormhole throat. Further, the left graph of Fig.\ref{fig2} indicates the asymptotic behavior of the shape function for different values of $\beta$. It shows that for increasing the larger values of the radial distance, the ratio $\frac{b(r)}{r}$ closes to zero, which guarantees the asymptotic behavior of the shape function. Also, in this case, we consider the wormhole throat $r_0=1$, and its graphical representation is shown in the right panel of Fig. \ref{fig2}.
 
\begin{figure*}
\centering
\includegraphics[width=14.5cm,height=5cm]{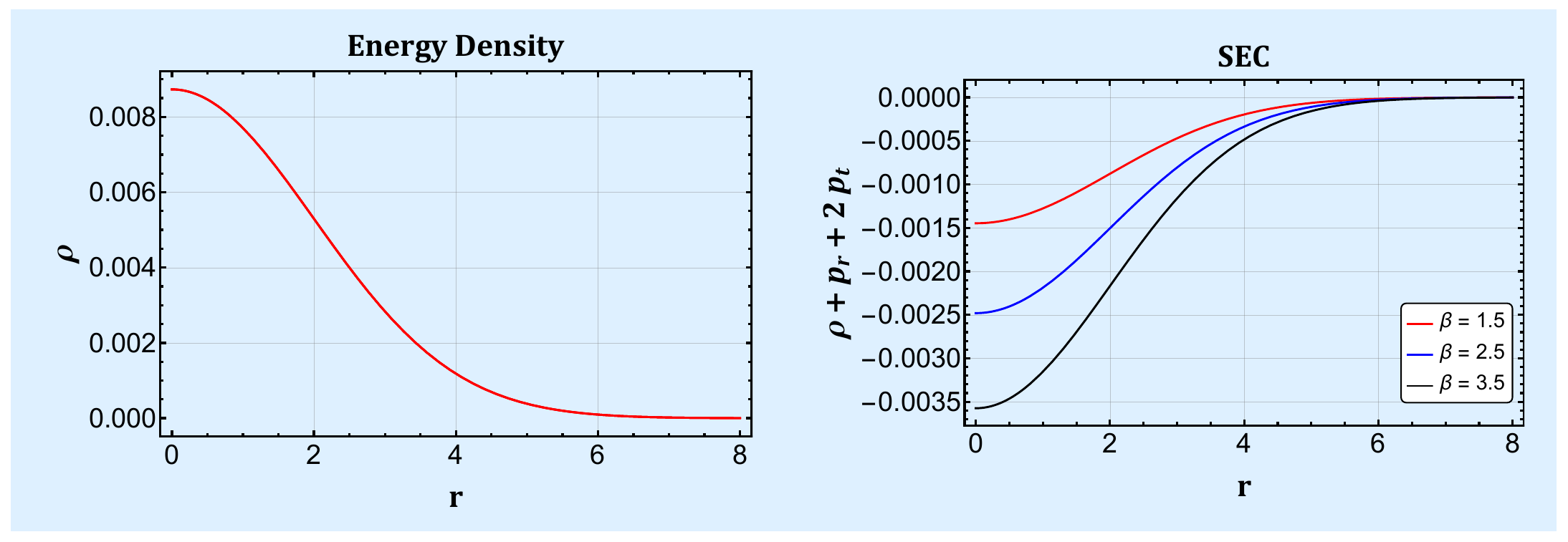}
\caption{Figure represents the behavior of energy density (\textit{left}) and SEC (\textit{right}) against radial coordinate $r$ for different values of $\beta$. Also, we consider $\Theta=2\, \text{and} \, M=1.1$. It is clear that $\rho$ shows positively decreasing behavior and $\rho+p_r+2p_t$  shows negative behavior. See the text for details.}
\label{fig3}
\end{figure*}

\begin{figure*}
\centering
\includegraphics[width=14.5cm,height=5cm]{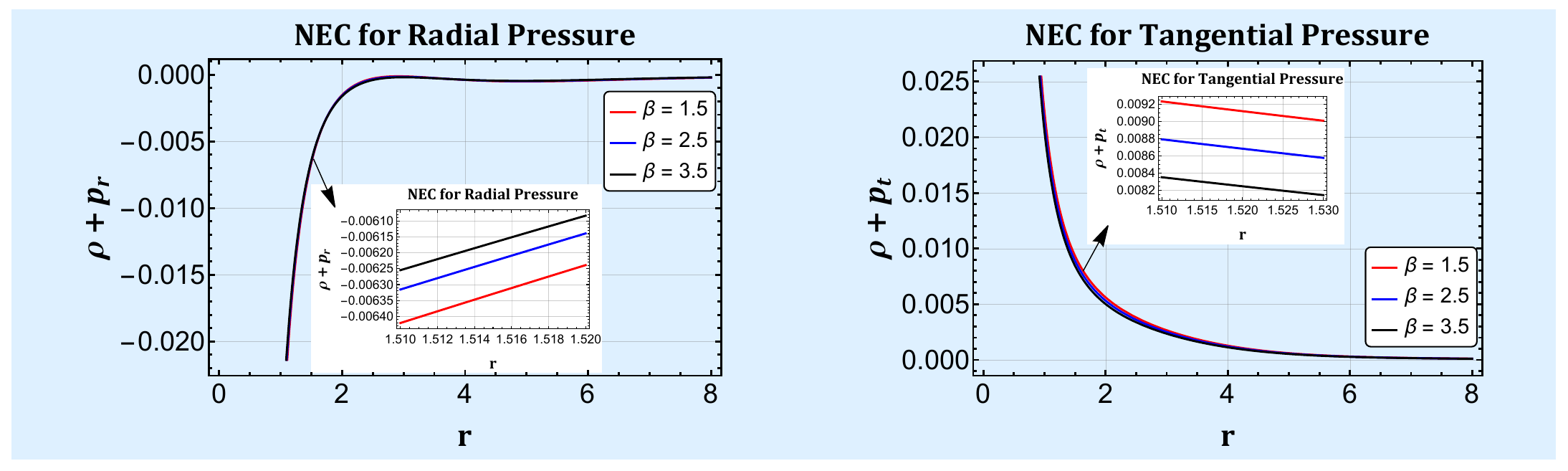}
\caption{Figure represents the behavior of NEC for radial (\textit{left}) and tangential (\textit{right}) pressure against radial coordinate $r$ for different values of $\beta$. Also, we consider $\alpha=1,\, \Theta=2,\, M=1.1,\, \text{and} \, r_0 = 1$. It is clear that $\rho+p_r$ shows negative behavior and $\rho+p_t$  shows positive behavior. See the text for details.}
\label{fig4}
\end{figure*}

Now to investigate the energy conditions, especially NEC and SEC, we substitute the obtained shape function \eqref{a3} in Eqs. \eqref{19} and \eqref{20}, and obtained the pressure components
\begin{multline}\label{26}
p_r = \frac{e^{-\frac{r^2+r_0^2}{4 \Theta }} }{\mathcal{K}_2\,(\beta +8 \pi )}\left[(\beta +8 \pi ) M e^{\frac{r_0^2}{4 \Theta }} \left(3 \sqrt{\pi } (4 \pi -\beta ) \Theta ^{3/2} 
\right.\right.\\\left.\left.
\times e^{\frac{r^2}{4 \Theta }} \mathcal{K}_3-\beta  r^3-3 (\beta -4 \pi ) \Theta  r\right)+\Theta  r_0 e^{\frac{r^2}{4 \Theta }} \left(-3 
\right.\right.\\\left.\left.
\times(4 \pi -\beta ) (\beta +8 \pi ) M-4 \pi ^{3/2} \alpha  (12 \pi -\beta ) \sqrt{\Theta } 
e^{\frac{r_0^2}{4 \Theta }}\right)\right],
\end{multline} 
\begin{multline}\label{27}
p_t = \frac{1}{4\mathcal{K}_2}\left[-6 \sqrt{\pi } (4 \pi -\beta ) \Theta ^{3/2} M \mathcal{K}_3+M e^{-\frac{r^2}{4 \Theta }}
\right. \\ \left.
\times \left(6 (\beta -4 \pi ) \Theta  r-(\beta +12 \pi ) r^3\right)+6 (4 \pi -\beta ) 
\right.\\\left.
\times \Theta  M r_0 e^{-\frac{r_0^2}{4 \Theta }}+\frac{8 \pi ^{3/2} \alpha  (12 \pi -\beta ) \Theta ^{3/2} r_0}{\beta +8 \pi }\right],
\end{multline}
where $\mathcal{K}_2=4 \pi ^{3/2} (12 \pi -\beta ) \Theta ^{3/2} r^3$ and \\
$\mathcal{K}_3=\left(\text{erf}\left(\frac{r_0}{2 \sqrt{\Theta }}\right)-\text{erf}\left(\frac{r}{2 \sqrt{\Theta }}\right)\right)$.\\
In this case, the NEC for radial and tangential pressures can be obtained as
\begin{multline}\label{28}
\rho + p_r = \frac{e^{-\frac{r^2+r_0^2}{4 \Theta }}}{2 \mathcal{K}_2 (\beta +8 \pi )}\left[3 (4 \pi -\beta ) (\beta +8 \pi ) M e^{\frac{r_0^2}{4 \Theta }} 
\right. \\ \left.
\times \left(2 \sqrt{\pi } \Theta ^{3/2} e^{\frac{r^2}{4 \Theta }} \mathcal{K}_3+r^3+2 \Theta  r\right)-2 \Theta  r_0 e^{\frac{r^2}{4 \Theta }} 
\right.\\\left.
\times \left(3 (4 \pi -\beta ) (\beta +8 \pi ) M+4 \pi ^{3/2} \alpha  (12 \pi -\beta ) \sqrt{\Theta } e^{\frac{r_0^2}{4 \Theta }}\right)\right],
\end{multline}
\begin{multline}\label{29}
\rho + p_t = \frac{1}{16 \pi ^{3/2} \Theta ^{3/2} r^3}\left[\frac{3 (4 \pi -\beta ) M e^{-\frac{r^2}{4 \Theta }}}{12 \pi -\beta } \left(-2 \sqrt{\pi } \Theta ^{3/2} e^{\frac{r^2}{4 \Theta }} \mathcal{K}_3
\right.\right.\\\left.\left.
+r^3-2 \Theta  r\right)+2 \Theta  r_0 
\left(\frac{4 \pi ^{3/2} \alpha  \sqrt{\Theta }}{\beta +8 \pi }+\frac{3 (4 \pi -\beta ) M e^{-\frac{r_0^2}{4 \Theta }}}{12 \pi -\beta }\right)\right],
\end{multline}
and at the throat i.e. at $r=r_0$, the above expressions for NEC reduce to
\begin{equation}
\rho+p_r\bigg\vert_{r=r_0}= \frac{3 (4 \pi -\beta ) M e^{-\frac{r_0^2}{4 \Theta }}}{8 \pi ^{3/2} (12 \pi -\beta ) \Theta ^{3/2}}-\frac{\alpha }{(\beta +8 \pi ) r_0^2}\,,
\end{equation}
\begin{equation}
\rho+p_t\bigg\vert_{r=r_0}=\frac{3 (4 \pi -\beta ) M e^{-\frac{r_0^2}{4 \Theta }}}{16 \pi ^{3/2} (12 \pi -\beta ) \Theta ^{3/2}}+\frac{\alpha }{2 (\beta +8 \pi ) r_0^2}\,.
\end{equation}
Also, the SEC for this case is given by
\begin{equation}\label{32}
\rho + p_r + 2 p_t = -\frac{\beta  M e^{-\frac{r^2}{4 \Theta }}}{\pi ^{3/2} (24 \pi -2 \beta ) \Theta ^{3/2}}.
\end{equation}
We note from the above expressions that $\beta\neq -8\pi\, \text{and}\, 12\pi$. Also, for $\beta=4\pi$, the effect of non-commutative geometry will no longer be available at the throat. By keeping these in mind, we have plotted the graphs for energy conditions in Figs. \ref{fig3}-\ref{fig4}. The graph for energy density versus $r$ is depicted in the left graph of Fig. \ref{fig3}, indicating the positively decreasing behavior in the entire spacetime. In contrast, the right graph shows the behavior of SEC for different values of $\beta$, representing the negative behavior. One can notice that the expression for SEC is independent of the model parameter $\alpha$, and hence for vanishing $\beta$, i.e., $\beta\rightarrow 0$, the Eq. \eqref{32} aligns with the result of \cite{Hassan1}. The left graph of Fig. \ref{fig4} shows the negative behavior for radial NEC, which leads to the violation of NEC. In contrast, the right graph confirms the positive behavior of tangential NEC in a decreasing way. Thus exotic matter may preserve wormhole solutions in the situation of non-metricity-based gravity with non-commutative geometry, which is the case for GR.

\subsection{Lorentzian distribution}
In this subsection,  we will discuss the wormhole solution under non-commutative geometry using Lorentzian distribution. The energy density for this case is given by \cite{P. Nicoloni,Mehdipour11}
 \begin{equation}\label{22}
 \rho =\frac{\sqrt{\Theta } M}{\pi ^2 \left(\Theta +r^2\right)^2}\,,
 \end{equation}
 where $M$ is the total mass and $\Theta$is the non-commutativity parameter.\\

\begin{figure*}
\centering
\includegraphics[width=14.5cm,height=6cm]{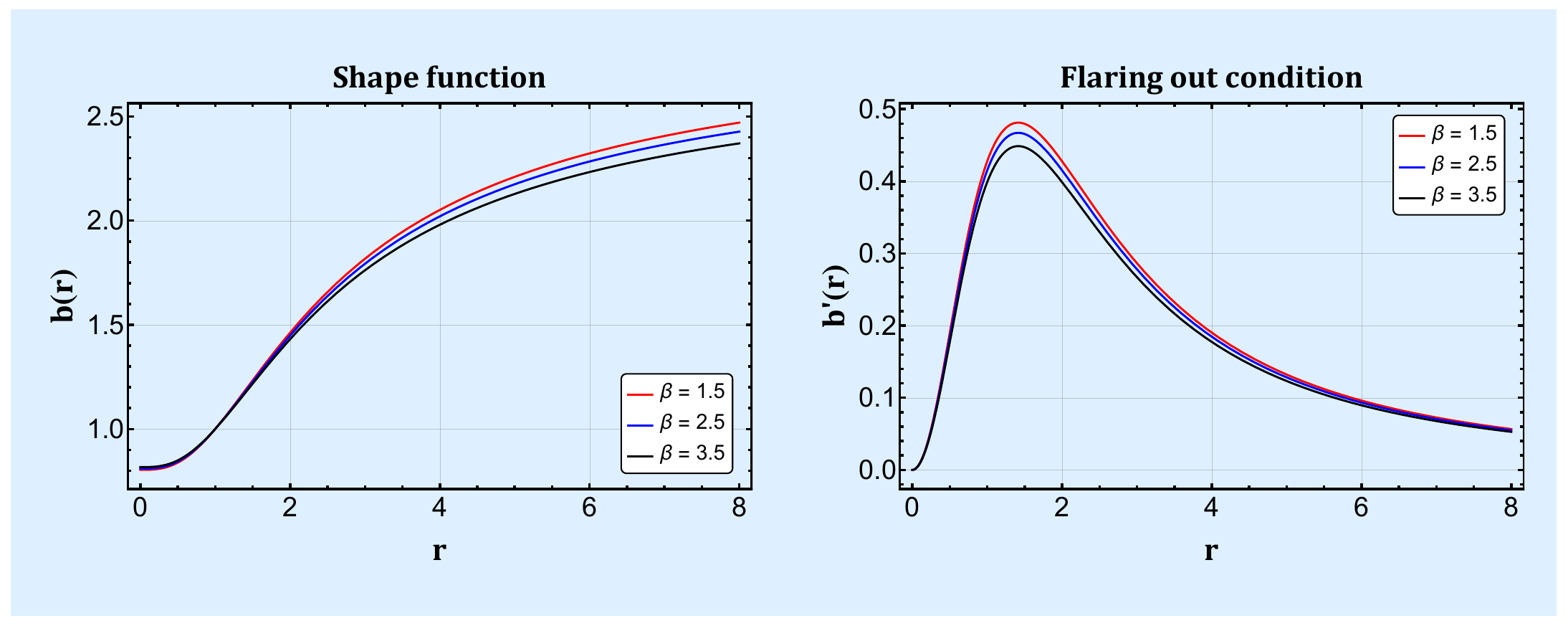}
\caption{Figure represents the behavior of shape function (\textit{left}) and flare-out condition (\textit{right}) against radial coordinate $r$ for different values of $\beta$. Also, we consider $\alpha=1,\, \Theta=2,\, M=1.1,\, \text{and} \, r_0 = 1$. It is clear that $b(r)$ shows a positively increasing behavior and $b'(r)$ satisfies the flaring out condition at the throat. See the text for details.}
\label{fig7}
\end{figure*}

\begin{figure*}
\centering
\includegraphics[width=14.5cm,height=6cm]{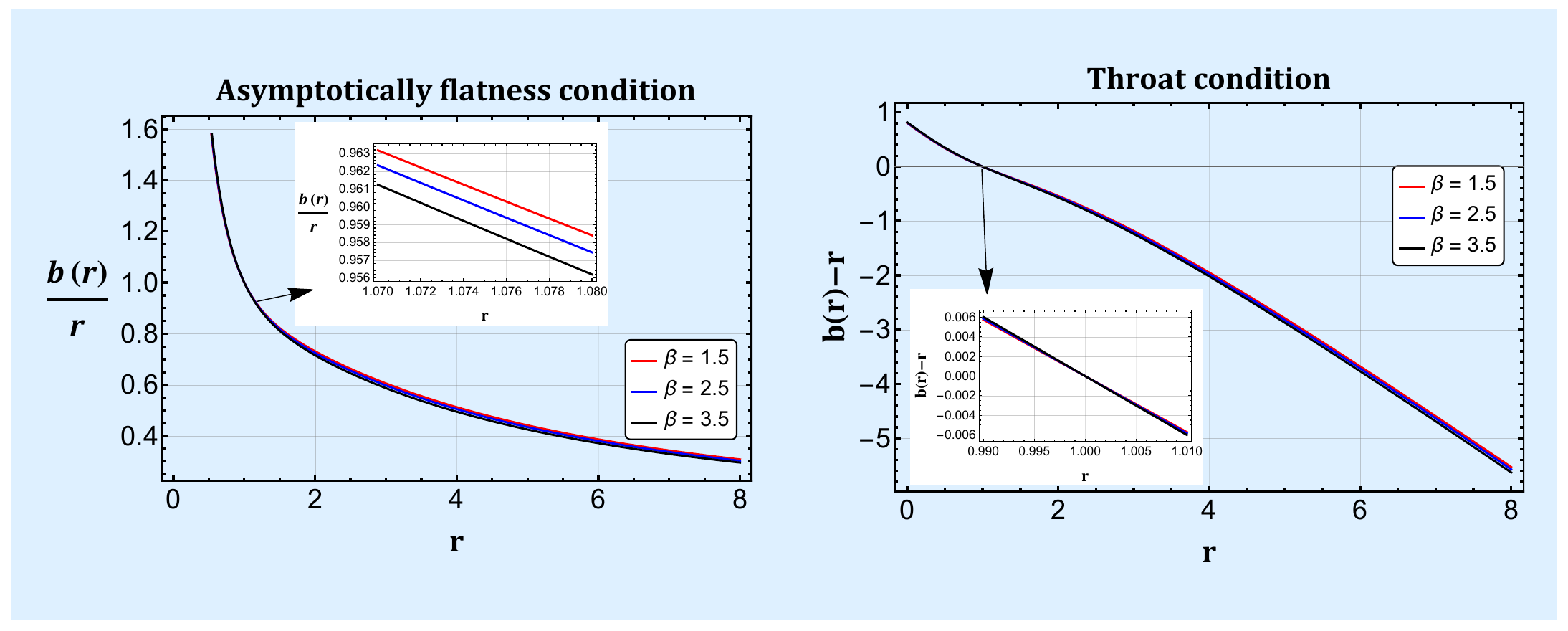}
\caption{Figure represents the behavior of asymptotically flatness condition (\textit{left}) and throat condition (\textit{right}) against radial coordinate $r$ for different values of $\beta$. Also, we consider $\alpha=1,\, \Theta=2,\, M=1.1,\, \text{and} \, r_0 = 1$. It is clear that $\frac{b(r)}{r}\rightarrow 0$ as $r\rightarrow \infty$ and $b(r)-r$  satisfies the throat condition. See the text for details.}
\label{fig8}
\end{figure*}

Now, By comparing the Eqs. \eqref{18} and \eqref{22} , the differential equation under Lorentzian distribution is given by 
 \begin{equation}\label{33}
 \frac{\alpha  (12 \pi -\beta ) b'(r)}{3 (4 \pi -\beta ) (\beta +8 \pi ) r^2}=\frac{\sqrt{\Theta } M}{\pi ^2 \left(\Theta +r^2\right)^2}\,.
 \end{equation}
 On integrating the above eq. \eqref{33} for the shape function $b(r)$, we get
 \begin{equation}\label{4b1}
b(r)=\frac{\mathcal{M}_1}{\left(\Theta +r^2\right)}  \left(\left(\Theta +r^2\right) \tan ^{-1}\left(\frac{r}{\sqrt{\Theta }}\right)-\sqrt{\Theta } r\right)+c_2\,,
 \end{equation}
where $c_2$ represents the constant of integration and $\mathcal{M}_1=\frac{3 (4 \pi -\beta ) (\beta +8 \pi ) M}{2 \pi ^2 \alpha  (12 \pi -\beta )}$. Now we impose the throat condition to the above equation and obtain $c_2$ as
\begin{equation}
c_2=r_0-\frac{\mathcal{M}_1}{\left(\Theta +r_0^2\right)}\left(\left(\Theta +r_0^2\right) \tan ^{-1}\left(\frac{r_0}{\sqrt{\Theta }}\right)-\sqrt{\Theta } r_0\right).
\end{equation}
Substituting the value of $c_2$ in Eq. \eqref{4b1}, we get
 \begin{multline}\label{34}
 b(r) = r_0+ \frac{\mathcal{M}_1}{\left(\Theta +r^2\right) \left(\Theta +r_0^2\right)}\left(\left(\Theta +r^2\right) \left(\Theta +r_0^2\right) 
 \right. \\ \left.
\mathcal{M}_2 +\sqrt{\Theta } (r-r_0) (r r_0-\Theta )\right).
 \end{multline}
where $\mathcal{M}_2=\left(\tan ^{-1}\left(\frac{r}{\sqrt{\Theta }}\right)-\tan ^{-1}\left(\frac{r_0}{\sqrt{\Theta }}\right)\right)$.\\
For Lorentzian distribution, the graphical behavior of the obtained shape function \eqref{34} and its related properties are shown in Figs. \ref{fig7} and \ref{fig8}. The left graph of Fig. \ref{fig7} indicates the positively increasing behavior of the shape function, while the right graph confirms that the flare-out condition is satisfied at the throat. In this case, also, we noticed that, for increasing the model parameter $\beta$, the shape function behaves in a decreasing way. In the left curve of Fig. \ref{fig8}, we have plotted for $\frac{b(r)}{r}$ versus radial coordinate $r$ for different values of $\beta$. It is seen that as we increase the values of the radial distance $r$, the ratio $\frac{b(r)}{r}$ converges to zero, which guarantees the asymptotic behaviors of the shape function. Here also, we consider the throat radius $r_0=1$. Thus, we can conclude that our obtained shape function obeyed all the properties for its traversability.\\

\begin{figure*}
\centering
\includegraphics[width=14.5cm,height=5cm]{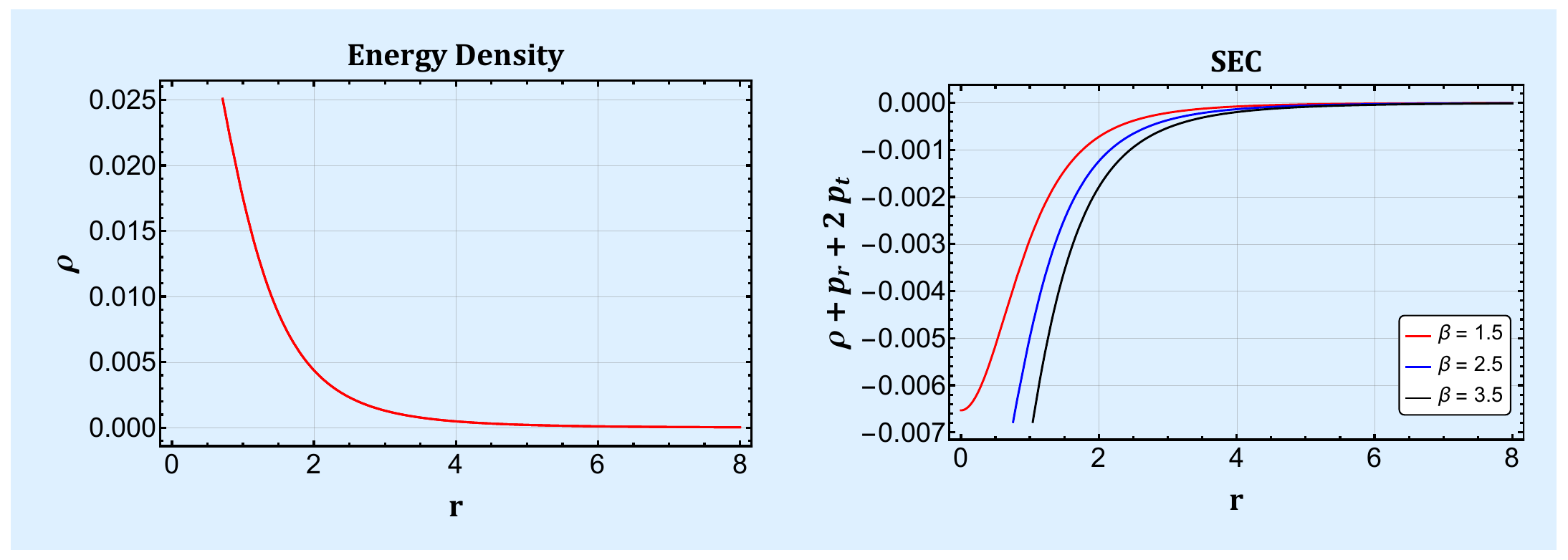}
\caption{Figure represents the behavior of energy density (\textit{left}) and SEC (\textit{right}) against radial coordinate $r$ for different values of $\beta$. Also, we consider $\Theta=2\, \text{and} \, M=1.1$. It is clear that $\rho$ shows positively decreasing behavior and $\rho+p_r+2p_t$  shows negative behavior. See the text for details.}
\label{fig9}
\end{figure*}

\begin{figure*}
\centering
\includegraphics[width=14.5cm,height=5cm]{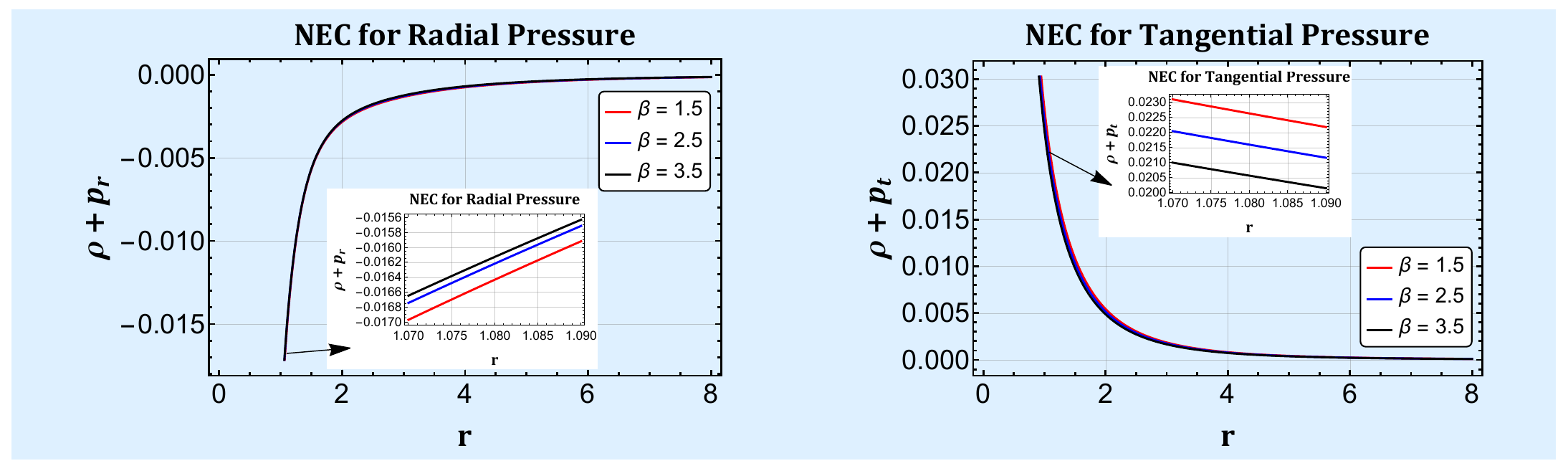}
\caption{Figure represents the behavior of NEC for radial (\textit{left}) and tangential (\textit{right}) pressure against radial coordinate $r$ for different values of $\beta$. Also, we consider $\alpha=1,\, \Theta=2,\, M=1.1,\, \text{and} \, r_0 = 1$. It is clear that $\rho+p_r$ shows negative behavior and $\rho+p_t$  shows positive behavior. See the text for details.}
\label{fig10}
\end{figure*}

For investigating the behavior of energy conditions under the Lorentzian distribution, we will use the shape function \eqref{34} in the Eqs. \eqref{18}-\eqref{20}. The obtained pressure components are
\begin{multline}\label{35}
p_r =\frac{1}{2 r^3}\left[\frac{M}{\pi ^2 (12 \pi -\beta ) \left(\Theta +r^2\right)^2} \left(\sqrt{\Theta } r \left(12 \pi  \left(\Theta +r^2\right)
\right.\right.\right.\\ \left.\left.\left.
-\beta  \left(3 \Theta +7 r^2\right)\right)-3 (4 \pi -\beta ) \left(\Theta +r^2\right)^2 \mathcal{M}_2\right) +r_0 
\right.\\ \left.
\times \left(\frac{3 (\beta -4 \pi ) \sqrt{\Theta } M}{\pi ^2 (12 \pi -\beta ) \left(\Theta +r_0^2\right)}-\frac{2 \alpha }{\beta +8 \pi }\right)\right]\,,
\end{multline} 
\begin{multline}\label{36}
p_t =-\frac{1}{\mathcal{M}_3}\left[\left(\Theta +r^2\right) \left(3 (4 \pi -\beta ) (\beta +8 \pi ) M \left(\Theta +r_0^2\right) 
\right.\right.\\\left.\left.
\times \left(\sqrt{\Theta } r-\left(\Theta +r^2\right) \tan ^{-1}\left(\frac{r}{\sqrt{\Theta }}\right)\right)-3 (4 \pi -\beta ) (\beta +8 \pi )
\right.\right.\\\left.\left.
 \times M\left(\Theta +r^2\right) \left(\sqrt{\Theta } r_0-\left(\Theta +r_0^2\right) \tan ^{-1}\left(\frac{r_0}{\sqrt{\Theta }}\right)\right)-2 \pi ^2 \alpha 
 \right.\right.\\\left.\left.
\times (12 \pi -\beta ) r_0 \left(\Theta +r^2\right) \left(\Theta +r_0^2\right)\right)
+2 (\beta +8 \pi ) (\beta +12 \pi ) 
\right. \\ \left.
\sqrt{\Theta } M r^3 \left(\Theta +r_0^2\right)\right],
\end{multline}
where $\mathcal{M}_3=4\pi^2(12\pi-\beta)(\beta +8\pi ) r^3\left(\Theta +r^2\right)^2 \\\left(\Theta+r_0^2\right)$.\\
At wormhole throat, the expressions for NEC can be read as
\begin{equation}
\rho+p_r\bigg\vert_{r=r_0}=\frac{3 (4 \pi -\beta ) \sqrt{\Theta } M}{\pi ^2 (12 \pi -\beta ) \left(\Theta +r_0^2\right)^2}-\frac{\alpha }{(\beta +8 \pi ) r_0^2}\,,
\end{equation}
\begin{equation}
\rho+p_t\bigg\vert_{r=r_0}=\frac{1}{2} \left(\frac{3 (4 \pi -\beta ) \sqrt{\Theta } M}{\pi ^2 (12 \pi -\beta ) \left(\Theta +r_0^2\right)^2}+\frac{\alpha }{(\beta +8 \pi ) r_0^2}\right).
\end{equation}
From the above expressions, it is clear that $\beta\neq -8\pi \,\text{and}\,12\pi$. We graphically explained the behavior of energy conditions in Figs. \ref{fig9} and \ref{fig10} for different values of $\beta$. As usual, energy density is positive in the entire spacetime. The Fig. \ref{fig10} represents the NEC versus radial distance $r$, graph indicating that $\rho+p_t$ is positive but decreasing behavior, whereas the radial NEC, i.e., $\rho+p_r$ shows the increasing negative behavior in the vicinity of the throat. SEC situation is the same as the Gaussian distribution (one may check the right graph of Fig. \ref{fig9}). This demonstrates that the NEC was violated in this instance, allowing the wormhole to exist. Consequently, the provided solutions are physically viable in both Gaussian and Lorentzian cases because they satisfy all the necessary characteristics of the shape function for the existence of the wormhole.\\

\section{Wormhole solutions with $f(Q,\,T)=Q+\,\lambda\,Q^2+\eta\,T$}
\label{sec5}
In this section, we assume a non-linear form of $f(Q,\,T)$ define by \cite{Tayde 3}
\begin{equation}
\label{41}
f(Q,\,T)=Q+\,\lambda\,Q^2+\eta\,T\,,
\end{equation}
 where $\lambda$ and $\eta$ are dimensionless model parameters. Note that, for $\lambda=0$, the above non-linear model will reduce to the linear model \eqref{17}; also, $\lambda=\eta=0$ gives equivalent to GR solution. This model successfully describes the de-Sitter type expansion of the Universe \cite{Loo}. Further, an investigation into wormholes was attempted; unfortunately, the authors did not find the exact solution because of the high non-linearity of the field equations \cite{Tayde 3}. Here, we are going to check the strength of the model in wormhole solutions under non-commutative geometries. Using the above model \eqref{41}, we can rewrite the components of the energy-momentum tensor from Eqs. (\ref{14}-\ref{16})
 \begin{multline}\label{42}
 \rho =\frac{1}{\mathcal{N}}\left(r b' \left(b \lambda  r b' (-11 b \eta +72 \pi  b+8 \eta  r-96 \pi  r) +2 b^2 
 \right.\right. \\ \left.\left.
 \lambda(15 b (\eta -8 \pi )+16 (12 \pi -\eta ) r)+4 (12 \pi -\eta ) r^3 (b-r)^2\right)
 \right.\\\left.
 +b^3 \lambda \left(3 b (88 \pi -9 \eta )+32 (\eta -12 \pi ) r\right)\right),
 \end{multline}
 \begin{multline}\label{43}
 p_r=\frac{1}{\mathcal{N}}\left(r b' \left(b \lambda  r b' (-13 b \eta +24 \pi  b+16 \eta  r)+2 b^2 \lambda  (3 b 
 \right.\right. \\ \left.\left.
 (3 \eta +8 \pi )-8 (\eta +12 \pi ) r)-8 \eta  r^3 (b-r)^2\right)
 +b^3 \lambda  (3 b 
 \right. \\ \left.
 (\eta -56 \pi )+8 (36 \pi -\eta ) r)-12 b (4 \pi -\eta ) r^3 (b-r)^2\right),
 \end{multline}
 \begin{multline}\label{44}
p_t= \frac{1}{\mathcal{N}}\left(r b' \left(b \lambda  r b' (4 (\eta +12 \pi ) r-b (\eta +24 \pi ))+2 b^2
\right.\right. \\ \left.\left.
\lambda  (9 b \eta +24 \pi  b-20 \eta  r-48 \pi  r)-2 (\eta +12 \pi ) r^3 
\right.\right. \\ \left.\left.
(b-r)^2\right)+b^3 \lambda  (-3 b (11 \eta +8 \pi )+52 \eta  r+48 \pi  r)
 \right. \\ \left.
  +6 b (4 \pi -\eta ) r^3(b-r)^2\right).
 \end{multline}
 where $\mathcal{N}=12 (4 \pi -\eta ) (\eta +8 \pi ) r^6 (b-r)^2$.\\
 The above equations are highly complicated; hence finding the exact solutions to these equations is extremely problematic. Thus, we will study the above-modified field equations with numerical approaches under both non-commutative distributions in the following consecutive subsections.
 
 \subsection{Gaussian distribution}
 \label{section4}
\begin{figure*}[h]
\centering
\includegraphics[width=14.5cm,height=6cm]{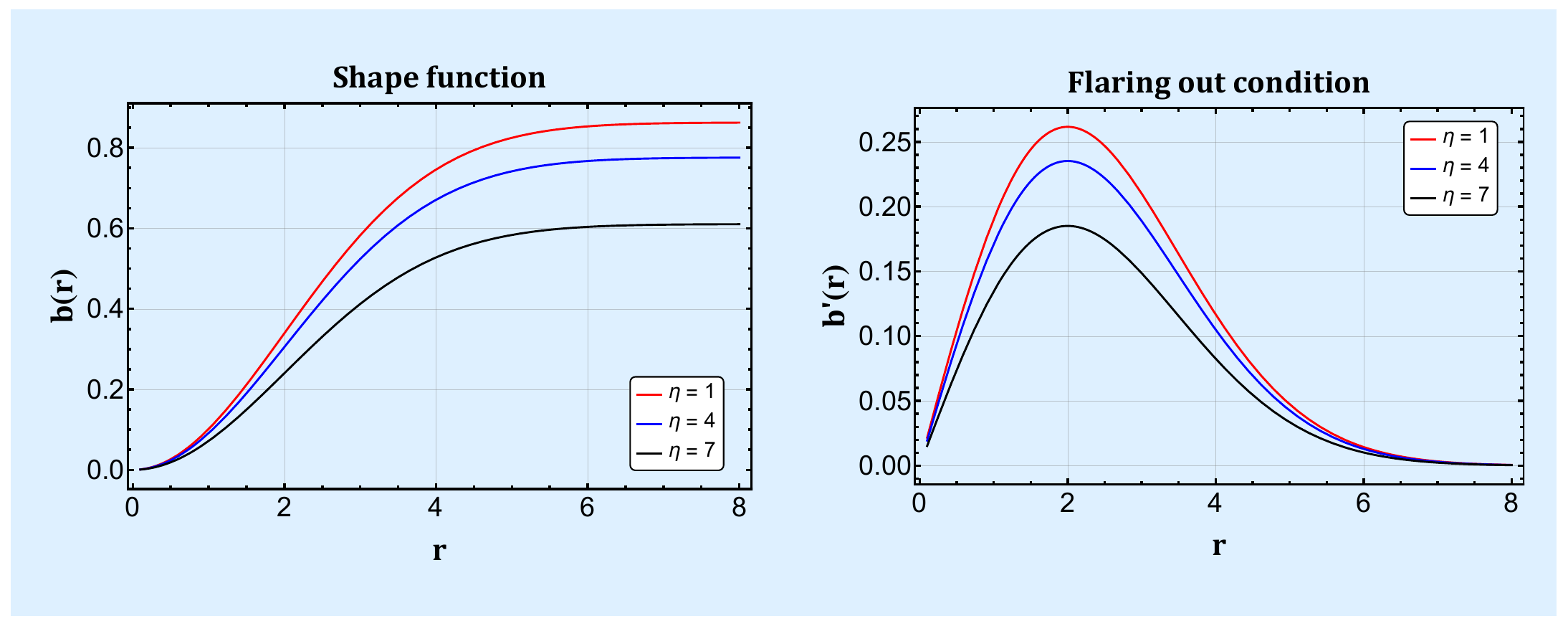}
\caption{Figure represents the behavior of shape function (\textit{left}) and flare-out condition (\textit{right}) against radial coordinate $r$ for different values of $\eta$. Also, we consider $\lambda=0.01,\, \Theta=2,\, \text{and} \, M=1.1$. It is clear that $b(r)$ shows a positively increasing behavior and $b'(r)$ satisfies the flaring out condition at the throat. See the text for details.}
\label{fig13}
\end{figure*}

\begin{figure*}[h]
\centering
\includegraphics[width=14.5cm,height=6cm]{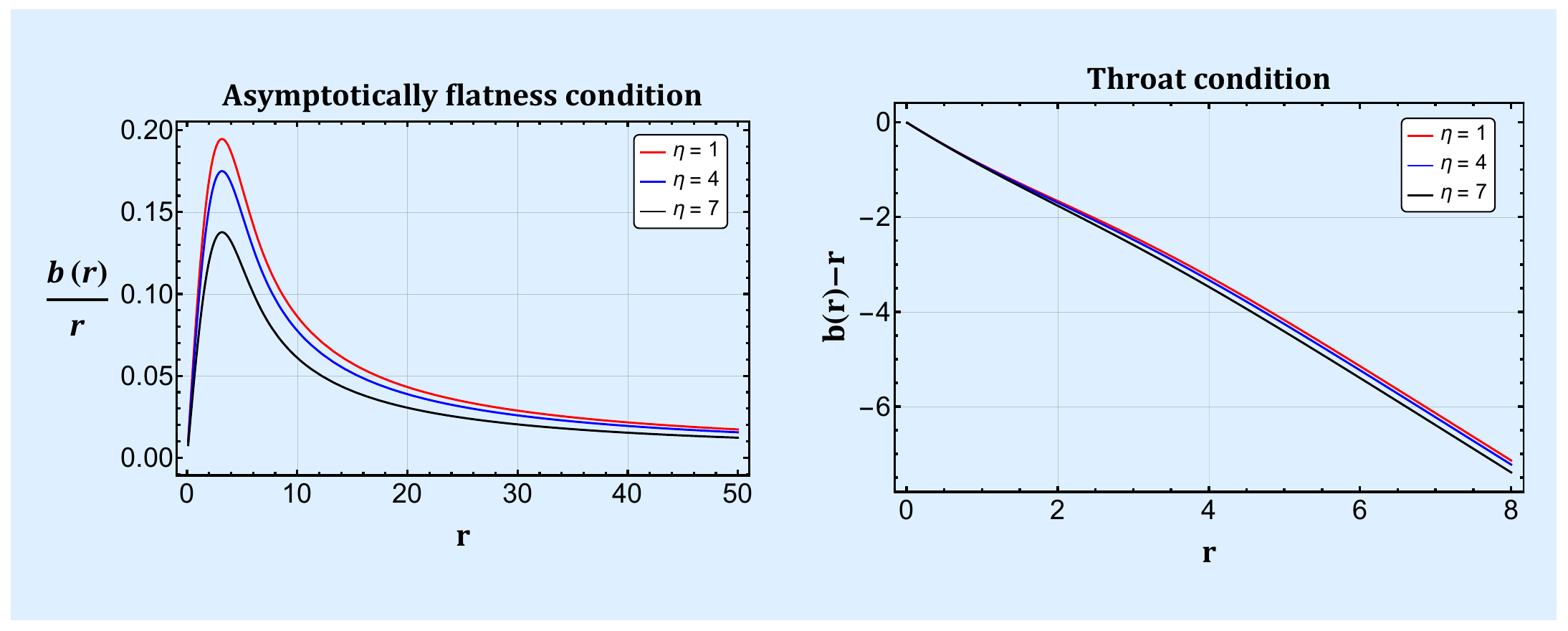}
\caption{Figure represents the behavior of asymptotically flatness condition (\textit{left}) and throat condition (\textit{right}) against radial coordinate $r$ for different values of $\eta$. Also, we consider $\lambda=0.01,\, \Theta=2,\, \text{and} \, M=1.1$. It is clear that $\frac{b(r)}{r}\rightarrow 0$ as $r\rightarrow \infty$ and $b(r)-r$ shows the location of the throat at 0.001 (approximately). See the text for details.}
\label{fig14}
\end{figure*}

\begin{figure*}
\centering
\includegraphics[width=14.5cm,height=5cm]{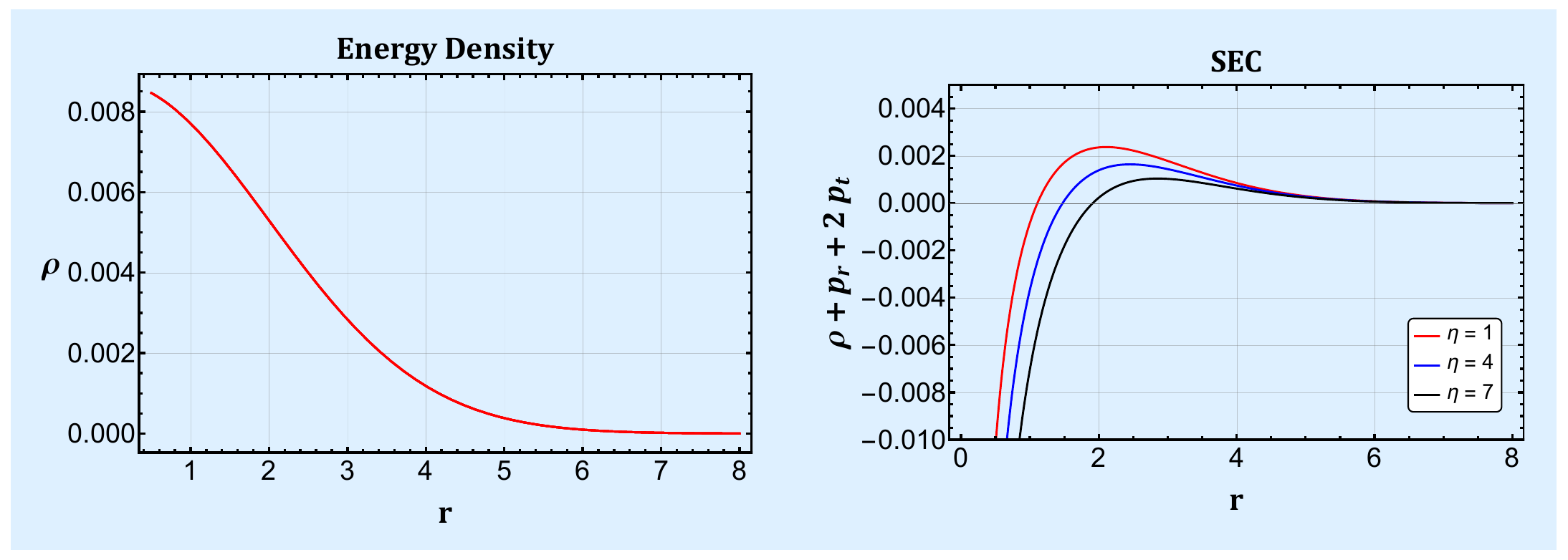}
\caption{Figure represents the behavior of energy density (\textit{left}) and SEC (\textit{right}) against radial coordinate $r$ for different values of $\eta$. Also, we consider $\lambda=0.01,\, \Theta=2,\, \text{and} \, M=1.1$. It is clear that $\rho$ shows positively decreasing behavior and $\rho+p_r+2p_t$  shows negative behavior at the throat. See the text for details.}
\label{fig15}
\end{figure*}

\begin{figure*}
\centering
\includegraphics[width=14.5cm,height=5cm]{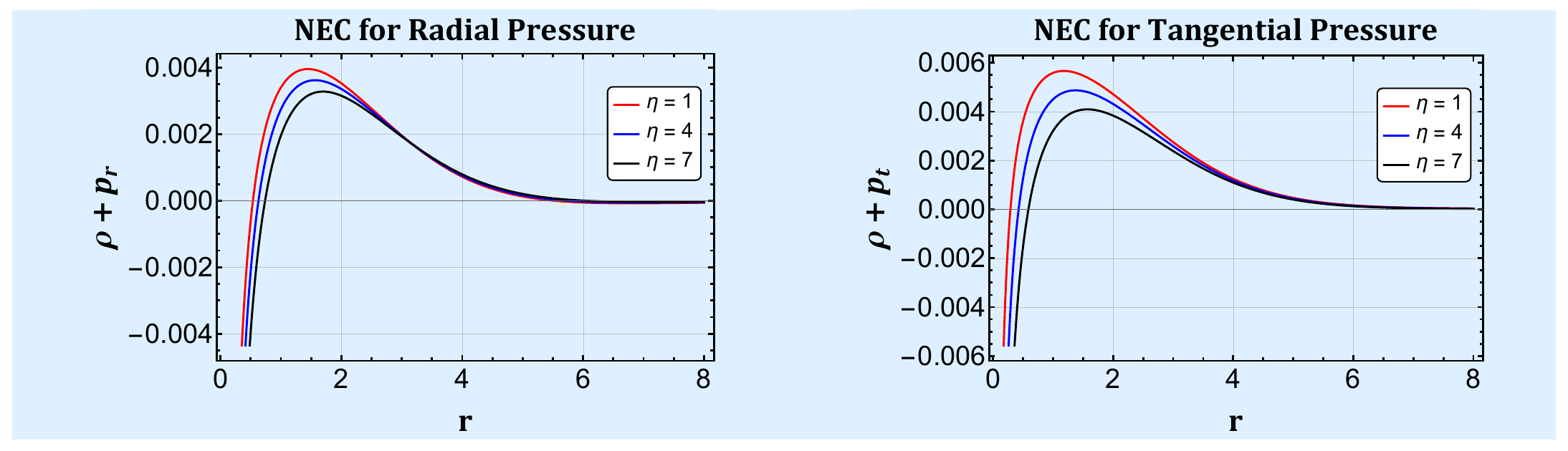}
\caption{Figure represents the behavior of NEC for radial (\textit{left}) and tangential (\textit{right}) pressure against radial coordinate $r$ for different values of $\eta$. Also, we consider $\lambda=0.01,\, \Theta=2,\, \text{and} \, M=1.1$. It is clear that NEC for both pressures shows negative behavior at the throat. See the text for details.}
\label{fig16}
\end{figure*}

We compare the Eq. \eqref{42} with the Gaussian distributed energy density \eqref{21} and obtain the following differential equation
 \begin{multline}\label{45}
     \frac{1}{\mathcal{N}}\left(r b' \left(b \lambda  r b' (-11 b \eta +72 \pi  b+8 \eta  r-96 \pi  r) +2 b^2 
 \right.\right. \\ \left.\left.
 \lambda(15 b (\eta -8 \pi )+16 (12 \pi -\eta ) r)+4 (12 \pi -\eta ) r^3 (b-r)^2\right)
 \right.\\\left.
 +b^3 \lambda \left(3 b (88 \pi -9 \eta )+32 (\eta -12 \pi ) r\right)\right)=\frac{M e^{-\frac{r^2}{4 \Theta }}}{8 \pi ^{3/2} \Theta ^{3/2}}\,,
 \end{multline}
 which is complicated to solve, and hence we solve it numerically for the shape function $b(r)$. Here, we consider the initial condition
 \begin{equation}\label{4a1}
   b(1\times 10^{-2})=1\times 10^{-5} . 
 \end{equation}
We find this initial condition in such a way that this condition satisfies all the important requirements of shape functions. We used the \textit{mathematica} tool with code \textit{NDSolve} and solved the Eq. \eqref{45} with the help of the initial condition \eqref{4a1} and discussed the necessary properties for the existence of wormhole structure graphically in Figs. \ref{fig13} and \ref{fig14}. Also, we have considered different possible values of the model parameters for the graphical views. The left plot of Fig. \ref{fig13} depicted the evolutions of the shape function, which indicates the positively increasing behavior, whereas for increasing the value of the model parameter $\eta$, the behavior of the shape function is decreasing. The derivative of the shape function versus $r$ is represented in the right plot of Fig. \ref{fig13}. We noticed that $b^{'}(r)<1$ at $r=r_0$, confirming that the flare-out condition is satisfied at the throat. Further, we plotted the graph for $\frac{b(r)}{r}$ with respect to $r$ for different positive values of model parameter $\eta$ in the left panel of Fig. \ref{fig14}, which shows that $\frac{b(r)}{r}\rightarrow 0$ as radial distance $r\rightarrow \infty$, for any positive values of $\eta$. However, the right plot corresponds to the function $b(r)-r$ versus $r$, which provides the location of the wormhole throat at $r_0 = 0.001$ (approximately). This kind of result can be found in \cite{a1,a2}. Furthermore, we checked the energy conditions for different values of $\eta$ shown in Figs. \ref{fig15} and \ref{fig16}. As usual, energy density shows positively decreasing behavior while SEC is violated in the vicinity of the throat. In addition, SEC decreases when the model parameter $\eta$ increases, and there is no effect of $\eta$ on energy density as it is not present in the equation of energy density for Gaussian distribution. Fig. \ref{fig16} indicates the violation of NEC near the throat for positive values of $\eta$, and far from the throat, NEC is satisfied. However, very far from the throat, NEC could be violated.

\subsection{Lorentzian distribution}
Similar to Gaussian distribution, we compare Eq. \eqref{42} with the Lorentzian distributed energy density \eqref{22}, and obtained a non-linear differential equation
\begin{multline}\label{46}
    \frac{1}{\mathcal{N}}\left(r b' \left(b \lambda  r b' (-11 b \eta +72 \pi  b+8 \eta  r-96 \pi  r) +2 b^2 
 \right.\right. \\ \left.\left.
 \lambda(15 b (\eta -8 \pi )+16 (12 \pi -\eta ) r)+4 (12 \pi -\eta ) r^3 (b-r)^2\right)
 \right.\\\left.
 +b^3 \lambda \left(3 b (88 \pi -9 \eta )+32 (\eta -12 \pi ) r\right)\right)=\frac{\sqrt{\Theta } M}{\pi ^2 \left(\Theta +r^2\right)^2},
\end{multline}
whose exact solution is also not possible. Thus we use the numerical technique to evaluate the possible form of the shape function. Here also, we consider the same initial condition \eqref{4a1} to solve the above Eq. \eqref{46}.\\
 \begin{figure*}[h]
\centering
\includegraphics[width=14.5cm,height=6cm]{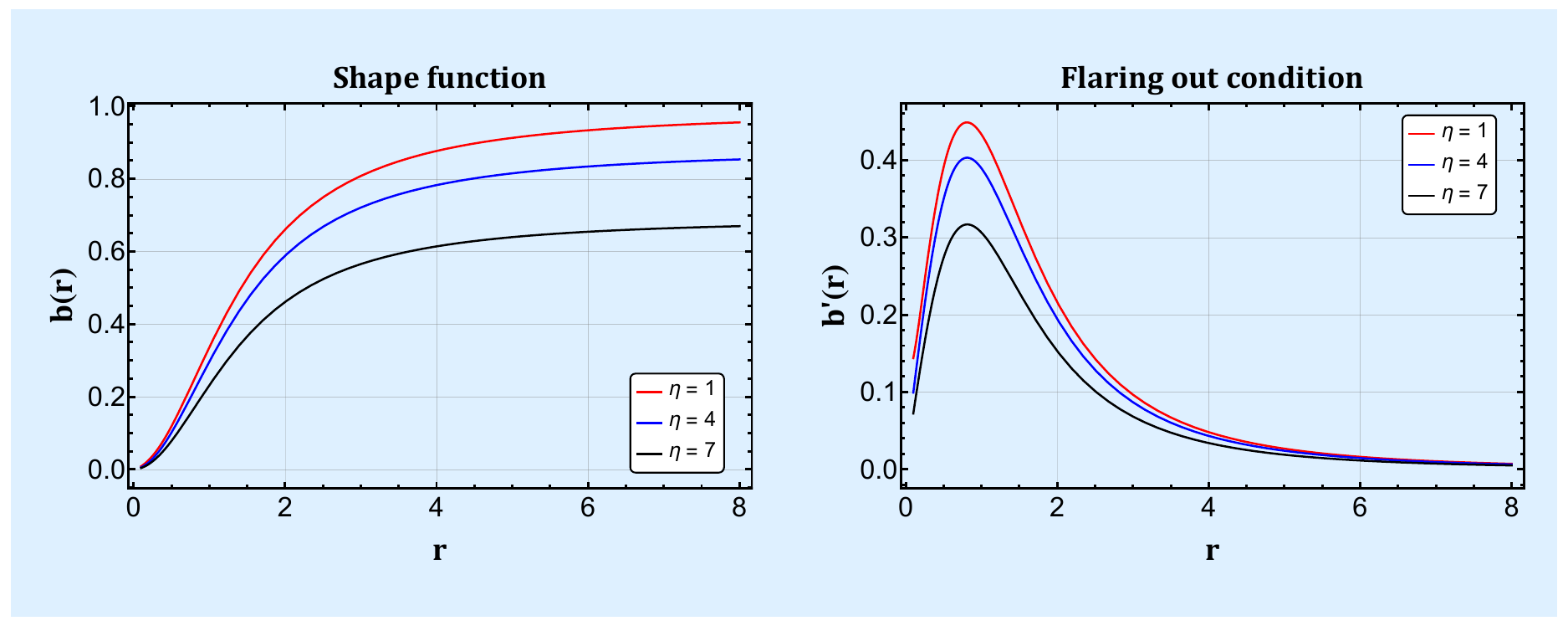}
\caption{Figure represents the behavior of shape function (\textit{left}) and flare-out condition (\textit{right}) against radial coordinate $r$ for different values of $\eta$. Also, we consider $\lambda=0.01,\, \Theta=2,\, \text{and} \, M=1.1$. It is clear that $b(r)$ shows a positively increasing behavior, and $b'(r)$ satisfies the flaring out condition at the throat. See the text for details.}
\label{fig19}
\end{figure*}
\begin{figure*}[h]
\centering
\includegraphics[width=14.5cm,height=6cm]{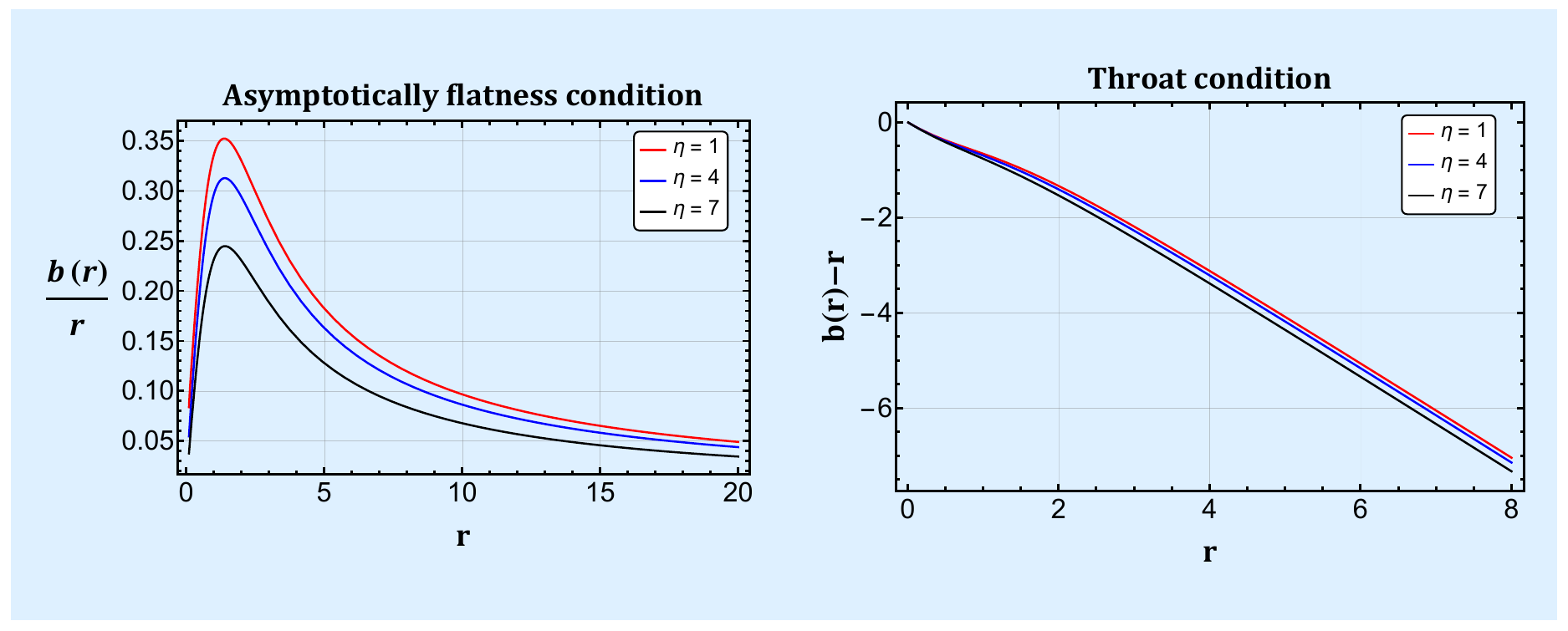}
\caption{Figure represents the behavior of asymptotically flatness condition (\textit{left}) and throat condition (\textit{right}) against radial coordinate $r$ for different values of $\eta$. Also, we consider $\lambda=0.01,\, \Theta=2,\, \text{and} \, M=1.1$. It is clear that $\frac{b(r)}{r}\rightarrow 0$ as $r\rightarrow \infty$, and $b(r)-r$ shows the location of the throat at 0.001 (approximately). See the text for details.}
\label{fig20}
\end{figure*}
\begin{figure*}
\centering
\includegraphics[width=14.5cm,height=5cm]{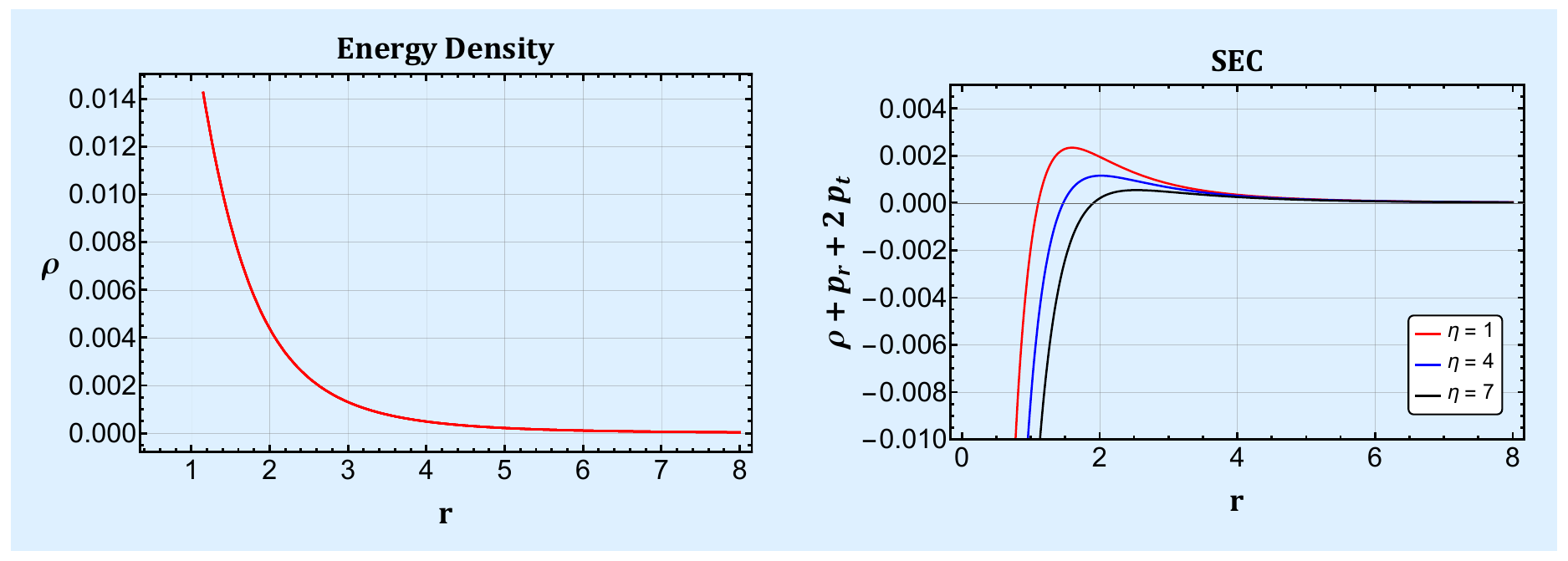}
\caption{Figure represents the behavior of energy density (\textit{left}) and SEC (\textit{right}) against radial coordinate $r$ for different values of $\eta$. Also, we consider $\lambda=0.01,\, \Theta=2,\, \text{and} \, M=1.1$. It is clear that $\rho$ shows positively decreasing behavior, and $\rho+p_r+2p_t$  shows negative behavior at the throat. See the text for details.}
\label{fig21}
\end{figure*}
\begin{figure*}
\centering
\includegraphics[width=14.5cm,height=5cm]{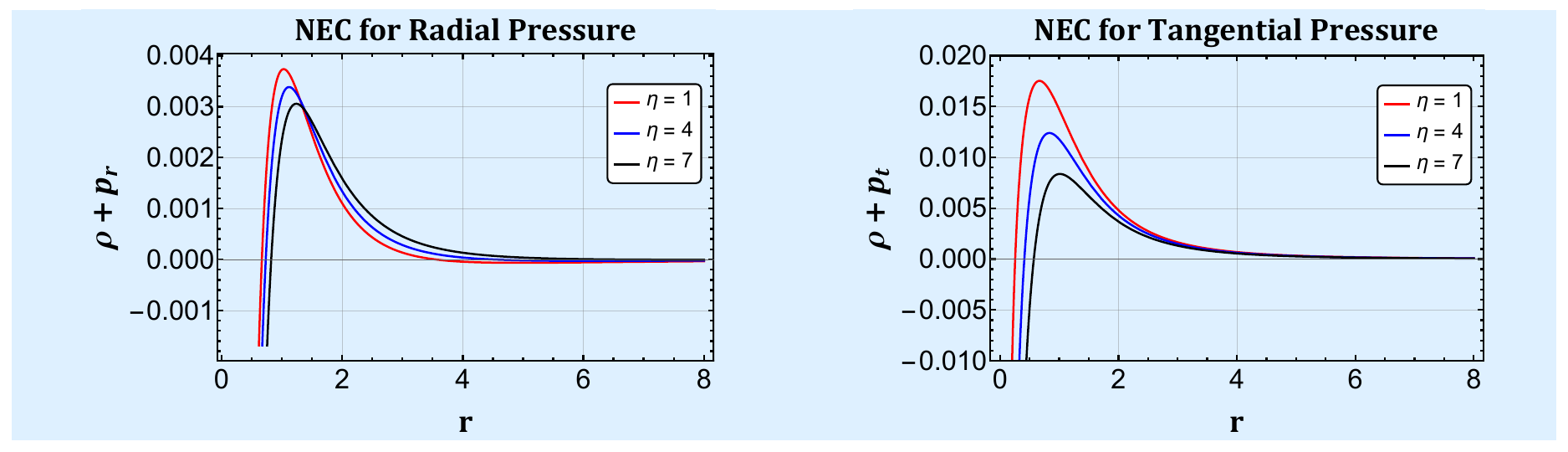}
\caption{Figure represents the behavior of NEC for radial (\textit{left}) and tangential (\textit{right}) pressure against radial coordinate $r$ for different values of $\eta$. Also, we consider $\lambda=0.01,\, \Theta=2,\, \text{and} \, M=1.1$. It is clear that NEC for both pressures shows negative behavior at the throat. See the text for details.}
\label{fig22}
\end{figure*}
Following the same procedure as the Gaussian distribution, we conducted a numerical solution of Eq. \eqref{46} and analyzed the corresponding properties of the wormhole structure. The initial condition \eqref{4a1} was utilized for this purpose also. The graphical representations in Figs. \ref{fig19} and \ref{fig20} were employed to illustrate our findings. Additionally, various values of the model parameters were considered in order to present a comprehensive view. The left plot of Fig. \ref{fig19} demonstrates the evolution of the shape function, indicating an increasing positive behavior. One can notice that as the model parameter $\eta$ increases, the shape function shows a decreasing trend. The right plot of Fig. \ref{fig19} represents the derivative of the shape function concerning $r$, confirming the fulfillment of the flare-out condition ($b'(r) < 1$) at the throat. In Fig. \ref{fig20}, the left panel exhibits the graph of $\frac{b(r)}{r}$ as a function of $r$ for various positive values of the model parameter $\eta$. It can be observed that as the radial distance $r$ tends towards infinity, $\frac{b(r)}{r}$ approaches zero. On the other hand, the right plot in Fig. \ref{fig20} shows the function $b(r)-r$ plotted against $r$, providing the location of the wormhole throat at approximately $r_0 \approx 0.001$. Furthermore, we investigated the energy conditions under Lorentzian distribution, depicted in Figs. \ref{fig21} and \ref{fig22}. As expected, the energy density exhibited a positive decreasing behavior, while the SEC was violated in the vicinity of the throat. Moreover, with an increase in the model parameter $\eta$, the violation of SEC became more prominent. Fig. \ref{fig22} confirms the violation of the NEC for both pressures at the throat for the chosen values of $\eta$. However, it is worth mentioning that NEC violations may still occur very far from the throat. During this study, we noticed that the effect of the model parameter and non-commutative parameter are responsible for the violation of energy conditions.
Thus, it can be concluded that the obtained numerical wormhole solutions for the non-linear $f(Q,\,T)$ model are physically interesting in both non-commutative distributions.

\section{Gravitational lensing}
\label{sec6}
In this segment, we shall use gravitational lensing phenomena to explore the possible detection of a traversable wormhole. The interest of Researchers in gravitational lensing, particularly in strong gravitational lensing, has increased over time after the works of Virbhadra et al. \cite{Virbhadra1, Virbhadra2, Virbhadra3}. Also, in \cite{Bozza2}, Bozza developed an analytic technique for calculating gravitational lensing in strong field limits for any spherically symmetric spacetime. Later this technique was used in many studies such as in \cite{Quinet, Tejeiro}. This motivates us to implement this method in this recent study. For this purpose, we consider the spherically symmetric line element involving the radius in Schwarzschild units, i.e., $x = \frac{r}{2\mathcal{M}}$ given by
\begin{equation}\label{5a}
ds^2=A(x)dt^2-B(x)dr^2-C(x)\,\left(d\theta^2+\sin^2\theta\,d\Phi^2\right)\,.
\end{equation}
Here $\bar{x} = \frac{\bar{r}}{2\mathcal{M}}$ denotes the closest path of the light ray. We will also take into account the exact shape function for this simple linear $f(Q,\,T)=\alpha\,Q+\beta\,T$ model given by equation \eqref{4b1}. Here, one can clearly observe that the model parameters $\alpha\neq0$ and $\beta\neq 4\pi, -8\pi,\,\, \text{and}\,\, 12\pi$ for the shape function \eqref{4b1} to exist. In essence, we take into account the wormhole metric \eqref{10}, where $e^{2\phi(r)}=\left(\frac{r}{b_0}\right)^m$, $b_0$ is an integration constant, and $m=2(v^\Phi)^2$ is the rotational velocity. It is noted in \cite{K. K. Nandi, Tejeiro, Kuhfittig2, Kuhfittig3} that $m \approx 0.000001$, a very tiny number (nearly zero), renders the redshift function constant (as we assumed in previous sections). The following connections are produced by comparing these metrics:
\begin{equation}
    A(x)=\left(\frac{r}{b_0}\right)^m;\quad B(x)= \left(1-\frac{b(r)}{r}\right)^{-1}; \quad C(x)=r^2\,.
\end{equation}
It is discussed in Ref. \cite{Tejeiro}, the deflection angle $\alpha(\bar{x})$ consists of the sum of two terms
\begin{equation}
\alpha(\bar{x})=\alpha_e\,+I(\bar{x}),
\end{equation}
where 
\begin{equation}
\alpha_e=-2 \ln \left(\frac{2a}{3}-1\right)-0.8056\,, 
\end{equation}
is due to the external Schwarzschild metric outside the wormhole's mouth $r=a$ and $I(\bar{x})$ is the contribution from the internal metric, which is given by
\begin{equation}
    I(\bar{x}) = 2\bigintss_{\bar{x}}^{\infty} \frac{\sqrt{B(x)}}{\sqrt{C(x)}\sqrt{\frac{C(x)A(\bar{x})}{C(\bar{x})A(x)}-1}} \,\,dx.
\end{equation}
We can simplify the above relation for the line element \eqref{5a} with the shape function \eqref{34}
\begin{equation}
    I(\bar{x}) = \int_{\bar{x}}^{a} R(x) dx\,,
\end{equation}
where 
\begin{widetext}
\begin{equation}
R(x)=\frac{2}{\sqrt{x^2 \left[1-\frac{1}{x}\Biggl\{\frac{\mathcal{L} \left(\left(\Theta +(2\mathcal{M})^2 x^2\right) \tan ^{-1}\left(\frac{2\mathcal{M}}{\sqrt{\Theta }}x\right)-2\mathcal{M}\sqrt{\Theta } x\right)}{ \left(\Theta +(2\mathcal{M})^2 x^2\right)}-\frac{\mathcal{L} \left(\left(\Theta +(2\mathcal{M})^2 x_0^2\right) \tan ^{-1}\left(\frac{2\mathcal{M}}{\sqrt{\Theta }} x_0\right)-2\mathcal{M} \sqrt{\Theta } x_0\right)}{ \left(\Theta +(2\mathcal{M})^2 x_0^2\right)}+x_0\Biggl\}\right]\left(\frac{x^{2-m}}{{\bar{x}}^{2-m}}-1\right)}}\,,
\end{equation}
\end{widetext}
and $\mathcal{L}=\frac{3 (4 \pi -\beta ) (\beta +8 \pi ) M}{4\mathcal{M} \pi ^2 \alpha  (12 \pi -\beta )
}$. To see where this integral diverges, we make the change of variable $y = \frac{x}{\bar{x}}$, so that
$x = \bar{x}y$ and $ x_0= \bar{x}y_0$:
\begin{equation}
     I(\bar{x}) = \int_{1}^{\frac{a}{\bar{x}}} \frac{2}{\sqrt{H(y)}} dy\,,
\end{equation}
where
\begin{widetext}
\begin{multline}
H(y)=\left[1-\frac{1}{\bar{x}y}\Biggl\{\frac{\mathcal{L} \left(\left(\Theta +(2\mathcal{M})^2 (\bar{x}y)^2\right) \tan ^{-1}\left(\frac{2\mathcal{M}}{\sqrt{\Theta }}\bar{x}y\right)-2\mathcal{M}\sqrt{\Theta } \bar{x}y\right)}{ \left(\Theta +(2\mathcal{M})^2 (\bar{x}y)^2\right)}-
\right. \\ \left.
\frac{\mathcal{L} \left(\left(\Theta +(2\mathcal{M})^2 (\bar{x}y_0)^2\right) \tan ^{-1}\left(\frac{2\mathcal{M}}{\sqrt{\Theta }} \bar{x}y_0 \right)-2\mathcal{M} \sqrt{\Theta } \bar{x}y_0 \right)}{ \left(\Theta +(2\mathcal{M})^2 (\bar{x}y_0)^2\right)}+\bar{x}y_0\Biggl\}\right]\left(y^{4-m}-y^2\right)\,.
\end{multline}
\end{widetext}

Letting $H(y)=g(y)\,f(y)$, where
\begin{widetext}
\begin{multline}
g(y)= 1-\frac{1}{\bar{x}y}\Biggl\{\frac{\mathcal{L} \left(\left(\Theta +(2\mathcal{M})^2 (\bar{x}y)^2\right) \tan ^{-1}\left(\frac{2\mathcal{M}}{\sqrt{\Theta }}\bar{x}y\right)-2\mathcal{M}\sqrt{\Theta } \bar{x}y\right)}{ \left(\Theta +(2\mathcal{M})^2 (\bar{x}y)^2\right)}-\\
\frac{\mathcal{L} \left(\left(\Theta +(2\mathcal{M})^2 (\bar{x}y_0)^2\right) \tan ^{-1}\left(\frac{2\mathcal{M}}{\sqrt{\Theta }} \bar{x}y_0 \right)-2\mathcal{M} \sqrt{\Theta } \bar{x}y_0 \right)}{ \left(\Theta +(2\mathcal{M})^2 (\bar{x}y_0)^2\right)}+\bar{x}y_0\Biggl\}\,,
\end{multline}
\end{widetext}
and
\begin{equation}
    f(y)= \left(y^{4-m}-y^2\right)\,.
\end{equation}
Taylor's series can be used to expand the function $H(y)$ around $y = 1$ as follows:
\begin{multline}
    H(y) = (2-m)g(1)(y-1)+\left[\frac{1}{2}(5-m)(2-m)g(1)+
    \right. \\ \left.
    (2-m)g'(1)\right](y-1)^2+ \text{higher powers}.
\end{multline}
Here we reduce the Taylor expansion to the second order. Note that the leading term in the aforementioned equation indicates whether the integral $I(\bar{x})$ converges or diverges. According to the Ref. \cite{Kuhfittig2}, the integral converges due to the leading term $(y-1)^{\frac{1}{2}}$ where $g(1)\neq 0$ whereas if $g(1)= 0$, the integral diverges because the second term leads to $\ln (y-1)$. If we choose the wormhole throat to be the light ray's closest approach, which is $\bar{r} = r_0$, then naturally $y_0 = \frac{x_0}{\bar{x}} $ and consequently $y_0 = 1$. Thus, $g(y)$ is reduces to
\begin{widetext}
\begin{multline}
g(y)= 1-\frac{1}{\bar{x}y}\Biggl\{\frac{\mathcal{L} \left(\left(\Theta +(2\mathcal{M})^2 (\bar{x}y)^2\right) \tan ^{-1}\left(\frac{2\mathcal{M}}{\sqrt{\Theta }}\bar{x}y\right)-2\mathcal{M}\sqrt{\Theta } \bar{x}y\right)}{ \left(\Theta +(2\mathcal{M})^2 (\bar{x}y)^2\right)}-\\
\frac{\mathcal{L} \left(\left(\Theta +(2\mathcal{M})^2 (\bar{x})^2\right) \tan ^{-1}\left(\frac{2\mathcal{M}}{\sqrt{\Theta }} \bar{x} \right)-2\mathcal{M} \sqrt{\Theta } \bar{x} \right)}{ \left(\Theta +(2\mathcal{M})^2 (\bar{x})^2\right)}+\bar{x}\Biggl\}.
\end{multline}
\end{widetext}
It is simple to confirm from the above equation that $g(1) = 0$, and hence the integral diverges. Thus, a photon sphere can be found at the throat, and such a photon sphere can be detected.

\section{Conclusions}
\label{sec7}
 In GR, the existence of wormhole solutions with some exotic matter always has curious researchers. One of the necessary conditions for wormhole formation is the availability of exotic matter because it results in NEC violation and hence allows the wormhole to exist. This violation is necessary for the theoretical concept of wormholes, as the negative energy associated with exotic matter stabilizes and maintains the wormhole structure. The Casimir effect is one such example where the vacuum fluctuations in quantum fields at ground stages create the attracting force between two parallel plates \cite{Jusufi11}. The formation of wormholes in modified theories has gained interest due to their incorporation of the effective energy-momentum tensor that disrespects the NEC without separately involving any exotic matter. Essentially, they provide alternative frameworks where the conventional energy conditions are relaxed or modified, enabling the exploration of phenomena such as wormholes without relying on exotic matter explicitly. In this paper, we have constructed wormhole solutions in the non-metricity-based modified gravity under the background of Gaussian and Lorentzian distributed non-commutative geometries. Also, we assume two functional forms of $f(Q,\,T)$ gravity such as linear $f(Q,\,T)=\alpha\,Q+\beta\,T$ and non-linear $f(Q,\,T)=Q+\lambda\,Q^2+\eta\,T$ models in our study. Our main theoretical observations are discussed below:
 
 At first, we discussed the wormhole solutions for the linear model with Gaussian and Lorentzian distributions. We compared the energy density of the modified gravity with the non-commutative distributed energy density and integrated it to obtain the shape function of the wormhole space-time metric. We plotted $b(r)$ versus radial distance to analyze the physical behavior of these acquired solutions. We noticed that the shape function respects the flare-out condition under asymptotic background under both distributions. Also, we have graphically shown the effect of the model parameter on shape functions. It was depicted that an increase in the model parameter $\beta$ decreases the shape function for both non-commutative distributions. However, in the throat, this impact is insignificant. Further, we checked both distributions' energy density, NEC, and SEC. It is noticed that energy density shows a positively decreasing behavior in the entire space-time for both distributions. SEC is violated in the vicinity of the throat. Radial NEC indicates negative behaviors, whereas tangential NEC shows positive behavior in the neighborhood of the throat. This demonstrates that the NEC was violated in this instance, allowing the wormhole to exist. Therefore the acquired solutions are viable, enabling the wormhole to exist in $f(Q,\,T)$ gravity under these non-commutative geometries.\\
 Later, we try to investigate the existence of wormhole solutions for the model $f(Q,\,T)=Q+\lambda\,Q^2+\eta\,T$ under the framework of both Gaussian and Lorentzian distributions. Due to the difficulty of analytic solutions in this instance, we were forced to solve the highly complex non-linear differential equations for $b(r)$ numerically. In this case, we consider the initial condition $b(1\times 10^{-2})=1\times 10^{-5}$ for this study. We have graphically represented the numerical solutions of shape functions and energy conditions for both distributions in Figs. (\ref{fig13}-\ref{fig22}). One can observe that the shape function shows positively increasing behavior; however, as we increase the values of the model parameter $\eta$, the shape function decreases. Also, one of the necessary conditions is the flare-out condition which is compatible with asymptotic background under both distributions. Further, we located the wormhole throat for these distributions at $r_0=0.001$ (approximately). This kind of result can be found in \cite{a1,a2}. Also, we noticed that energy density shows entirely positive behavior. SEC is disrespected near the throat; however, far from the throat, SEC will be validated. NEC shows negative behavior, and hence it is incompatible with this solution. Hence, all the requirements for the existence of a wormhole are satisfied for the preferred specific values of free parameters, and consequently, the acquired wormhole solutions are viable.\\
At last, we have examined the effects of gravitational lensing for the wormhole solutions under non-commutative background. In order to do this, we followed the approach given in Refs. \cite{Tejeiro,Kuhfittig2,Kuhfittig3}  and explored the convergence of deflection angle. This technique was first introduced by Bozza et al. \cite{Bozza1} to investigate the black hole physics in the strong field limit. Later, Bozza \cite{Bozza2} developed an analytical technique to obtain gravitational lensing for the generic spherically symmetric metric in the strong field limit. Also, in \cite{Iftikhar}, the authors studied the behavior of the deflection angle and location of the wormhole through its shape function. Following the same procedure, we have studied the convergence of the deflection angle using the exact solutions for the shape function. The obtained results show the divergence of the deflection angle of the outward light ray, which occurs exactly at the throat of the wormhole representing the correspondence of the surface to the photon sphere. Thus, the integral divergence of the deflection angle at the throat indicates the potential presence of a photon sphere. The detection of a photon sphere in close proximity to a wormhole throat would have considerable significance due to its implications. It would confirm the presence of a strong gravitational field and validate predictions related to wormholes. Moreover, from the perspective of observational astronomy, identifying such a phenomenon would present an opportunity to directly investigate and observe these enigmatic structures, contributing to our understanding of gravity and the inherent nature of wormholes.\\
To summarize, the $f(Q,T)$ gravity allows us to study wormhole solutions under the effect of non-commutative geometries that violates energy conditions in some regions of space-time. We first consider a linear model that provides analytical exact wormhole solutions for both distributions. The obtained shape functions obey the flare-out condition under the asymptotic flatness condition, which are the essential requirements for a traversable wormhole. Also, these kinds of solutions violate energy conditions, especially NEC and SEC in the vicinity of the throat. In addition, we numerically studied wormhole solutions for the non-linear model under both distributions. In this case, we found that all the energy conditions are disrespected only in some regions of the space-time. We also noticed the effect of the model parameter and the non-commutative parameter in the shape functions and energy conditions under both Gaussian and Lorentzian distributions. In a recent study, it was noticed that due to non-commutative geometry, asymptotically flatness behavior of the shape function could not be achieved in $f(Q)$ gravity with conformal symmetry \cite{Mustafa11}. Interestingly, in this study, we found that the asymptotic flatness behavior is achieved for both distributions. Since the main issue of wormhole geometry is the exotic matter and the modified gravity allows us to avoid or minimize the usage of the exotic matter. Thus, in that case, the model parameter and non-commutative geometry may become the source of violation of NEC in the symmetric teleparallel with matter coupling case. Similar behavior can also be found in the teleparallel \cite{R10} case. Note that the current investigation has been done under the constant redshift function. However, considering the constant redshift function may not be very realistic from a physics perspective. In that case, one can relax this restriction and study wormhole solutions for non-constant redshift functions in more general scenarios. Thus, it would be interesting to investigate these solutions for non-constant redshift functions. We leave it to a future study.

\acknowledgments  MT acknowledges University Grants Commission (UGC), New Delhi, India, for awarding National Fellowship for Scheduled Caste Students (UGC-Ref. No.: 201610123801). ZH acknowledges the Department of Science and Technology (DST), Government of India, New Delhi, for awarding a Senior Research Fellowship (File No. DST/INSPIRE Fellowship/2019/IF190911). PKS acknowledges National Board for Higher Mathematics (NBHM) under the Department of Atomic Energy (DAE), Govt. of India, for financial support to carry out the Research project No.: 02011/3/2022 NBHM(R.P.)/R\&D II/2152 Dt.14.02.2022 and Transilvania University of Brasov for Transilvania Fellowship for Visiting Professors. We are very much grateful to the honorable referees and to the editor for the illuminating suggestions that have significantly improved our work in terms
of research quality, and presentation.


\end{document}